  \documentclass{aa}

  \usepackage{graphicx}
  \usepackage{txfonts}
  \usepackage{bm}
  \usepackage[breaklinks=true]{hyperref}
  \usepackage{xcolor}
  \hypersetup{
      colorlinks,
      linkcolor={blue!50!black},
      citecolor={blue!50!black},
      urlcolor={blue!50!black}
  }
  \usepackage{natbib}
  \bibpunct{(}{)}{;}{a}{}{,}
  \usepackage{lscape}

  \usepackage{subcaption}
  \newcommand{\Msun}{\rm M_{\sun}}

  \newcommand{\lsim}{\mathrel{\hbox{\rlap{\lower.55ex\hbox{$\sim$}} \kern-.3em\raise.4ex\hbox{$<$}}}}
  \newcommand{\gsim}{\mathrel{\hbox{\rlap{\lower.55ex\hbox{$\sim$}} \kern-.3em\raise.4ex\hbox{$>$}}}}

  \newcommand{\TNGF}{TNG50}

  \defcitealias{rosasguevara2020}{RG20}
  \defcitealias{rosasguevara2022}{RG22}

\begin{document}

    \title{Bulgeless Evolution And the Rise of Discs (BEARD)}

    \subtitle{II. The role of mergers in shaping the Milky Way analogues in TNG50}
    \titlerunning{BEARD II: Milky Way analogues in TNG50}

 \author{Yetli Rosas-Guevara\inst{1}\fnmsep\inst{2}\thanks{email:yetli.rosas@uco.es}
          \and
          Jairo M\'endez-Abreu \inst{3}\fnmsep\inst{4}
          \and
          Adriana de Lorenzo-C\'aceres \inst{3}\fnmsep\inst{4}
          \and
          Salvador Cardona-Barrero \inst{3}\fnmsep\inst{4}
          \and
          Elena Arjona-G\'alvez\inst{3}\fnmsep\inst{4}
          \and
          Mario Chamorro Cazorla\inst{12}\fnmsep\inst{14}
          \and
          Enrico Maria Corsini\inst{6}\fnmsep\inst{7}
          \and
          Luca Costantin\inst{11}
          \and
          Virginia Cuomo\inst{13}
          \and
          Arianna Di Cintio\inst{3}\fnmsep\inst{4}
          \and
          David Fernandez\inst{9}\fnmsep\inst{15}\fnmsep\inst{16}
          \and
          Daniele Gasparri\inst{8}
          \and
          Carlos Marrero-de la Rosa\inst{3}\fnmsep\inst{4}
          \and
          Divakara Mayya\inst{9}
          \and
          Lorenzo Morelli \inst{8}
          \and
          Casiana Mu\~{n}oz-Tu\~{n}\'on\inst{3}\fnmsep\inst{4}
           \and
          Francesca Pinna\inst{3}\fnmsep\inst{4}
          \and
          Alessandro Pizzella\inst{6}\fnmsep\inst{7}
          \and
          Javier Rom\'an\inst{12}
          \and
          Daniel Rosa Gonzalez\inst{9}
          \and
          Rubén S\'anchez-Janssen\inst{4}\fnmsep\inst{5}
          \and
          Olga Vega\inst{9}
          \and
          Stefano Zarattini\inst{10}
          }

  \institute{Donostia International Physics Centre (DIPC), Paseo Manuel de Lardizabal 4, 20018 Donostia-San Sebastian, Spain.
            \and
            Departamento de F\'isica, Universidad de C\'ordoba, Campus Universitario de Rabanales, Ctra. N-IV Km. 396, E-14071 C\'ordoba, Spain
            \and
            Universidad de La Laguna, Departamento de Astrof\'isica. Avda. Astrof\'isico Francisco S\'anchez S/N. E-38200 San Crist\'obal de La Laguna, Spain.
            \and
            Instituto de Astrof\'isica de Canarias. C/ V\'ia L\'actea S/N, E-38205 San Crist\'obal de La Laguna, Spain.
            \and
             Isaac Newton Group of Telescopes, E-38700 Santa Cruz de La Palma, Canary Islands, Spain.
             \and
              Dipartimento di Fisica e Astronomia ``G. Galilei'', Universit\`a di Padova, vicolo dell'Osservatorio 3, I-35122 Padova, Italy.
              \and
              INAF-Osservatorio Astronomico di Padova, vicolo dell'Osservatorio 2, I-35122 Padova, Italy.
              \and
              Instituto de Astronomía y Ciencias Planetarias, Universidad de Atacama, Copayapu 485, Copiap\'o, Chile.
              \and
              Instituto Nacional de Astrof\'isica, Optica y Electr\'onica, Tonantzintla, 72840 Puebla, M\'exico.
              \and
              Centro de Estudios de F\'isica del Cosmos de Arag\'on (CEFCA), Plaza San Juan 1, 44001 Teruel, Spain.
              \and
              Centro de Astrobiolog\'ia, CSIC-INTA, Ctra. de Ajalvir km 4, Torrej\'on de Ardoz, E-28850, Madrid, Spain.
              \and
              Departamento de F\'isica de la Tierra y Astrof\'isica, Universidad Complutense de Madrid, 28040 Madrid, Spain.
              \and
              Departamento de Astronom\'ia, Universidad de La Serena, Av. Ra\'ul Bitr\'an 1305, La Serena, Chile.
              \and
              Instituto de F\'isica de Part\'iculas y del Cosmos (IPARCOS), Facultad de Ciencias F\'isicas, Universidad Complutense de Madrid, E-28040 Madrid, Spain.
              \and
              Planetarium La Ense\~{n}anza, Medell\'in, Antioquia, CP. 050022 Colombia.
              \and
              Canada–France–Hawaii Telescope, Kamuela, HI 96743, USA.
              }

    \date{Received September 15, 1996; accepted March 16, 1997}

  % \abstract{}{}{}{}{}
  % 5 {} token are mandatory

    \abstract{
 We study the formation and evolution of bulgeless galaxies within the Milky Way-Andromeda analogue sample of the TNG50 simulation. Through kinematic decomposition with \textsc{Mordor}, we identified bulgeless galaxies with a bulge-to-disc mass ratio of $B/D\leq0.08$, in line with the Bulgeless Evolution And the Rise of Discs (BEARD) survey and Milky Way constraints. We compared the identified bulgeless galaxies to those that are bulge-dominated, which have a bulge-to-disc mass ratio of $B/D>1$. We find that $74\%$ of bulgeless galaxies experience at least one major merger (stellar mass ratio $1:4$) throughout their lifespan. Bulgeless galaxies form later ($z_{50} \sim 0.7$) compared to bulge-dominated counterparts ($z_{50} \sim 1.2$). Bulgeless galaxies have lower-mass haloes and higher specific stellar angular momentum, which is compatible with Milky Way observations. However, specific star formation rates and hydrogen gas fractions are slightly higher than Milky Way observations. Our analysis of the redshift evolution of stellar components reveals that bulgeless galaxies have gradual disc growth with high thin disc-to-total mass ratios ($D/T>0.5$) since $z\sim1$ and minimal bulge growth ($B/T<0.1$) since $z\sim1.5$. In contrast, bulge-dominated galaxies have earlier disc formation, which is disrupted, resulting in higher morphology evolution. Bulgeless galaxies are more likely to undergo gas-rich, coplanar, and corotating mergers, promoting disc survival, compared to bulge-dominated galaxies that encounter a broader spectrum of mergers. We also observed differences in galaxy structure between bulgeless and bulge-dominated galaxies without major mergers, suggesting the relevance of early gas accretion and alignment. Bulgeless galaxies have younger stellar populations and more extended star formation histories than bulge-dominated galaxies, which rapidly quench and have older stellar populations. These findings elucidate the distinct merger-driven and secular pathways that give rise to Milky Way galaxies.}
    %{Bulgeless galaxies ...}
    %% results heading (mandatory)
    %{}



    \maketitle
  %
  %-------------------------------------------------------------------

\section{Introduction}

Bulgeless (BL) galaxies are defined as disc-dominated galaxies with little or no central bulge. Within the $\Lambda$CDM (Cold Dark Matter) paradigm, galaxies assemble hierarchically through frequent mergers \citep{white1978}, which are expected to disrupt discs and promote the formation of central bulges \citep{brooks2016}. Therefore a high fraction of BL or pure disc galaxies is unexpected. Observationally, \citet{kormendy2010} showed that nearly $50\%$ of massive disc galaxies within a sphere of $8$ Mpc centred on the Milky Way (MW) exhibit no kinematic or photometric signs of a merger-built bulge. A similar fraction was found by \citet{weinzirl2009} based on H-band imaging of massive spirals.

The formation and evolution of massive disc galaxies with negligible bulges remains an open question. Several scenarios have been proposed. One possibility is that these galaxies experience quiet merger histories, allowing gas to retain its angular momentum and form extended discs \citep[e.g.,][]{sotillo2022,rodriguez2025}. Internal processes, such as gravitational instabilities, may lead to the gradual formation of central structures \citep[e.g.,][]{elmegreen2004,gadotti2009,shen2010,sellwood2014}. In addition, gas accretion and minor interactions can also contribute to disc growth without contributing to the growth of the bulge \citep[e.g.,][]{bertola1996,coccato2013,pizzella2018}.

In particular, \cite*{shen2010} found from kinematic modelling of the MW bulge, no evidence of the bulge of being formed through mergers. Instead, the authors found that the MW bulge appears to be the result of bar-driven instabilities within the disc, suggesting that the MW may be considered a bulgeless galaxy. However, galaxy formation models predict that $\sim70\%$ of MW-mass haloes experience at least one merger with a mass ratio of $1:10$ or larger during their lifetime \citep[e.g.,][]{stewart2008}. Indeed, observations from the {\it Gaia} mission have revealed a population of stars with elongated orbits, interpreted as debris from the Gaia--Sausage--Enceladus merger \citep{belokurov2018}. This event, with a mass ratio of roughly $1:4$, likely heated the MW early disc, contributing to the formation of its thick disc approximately $10$ Gyr ago \citep{helmi2018,belokurov2018}.

The role of mergers in the formation of galaxies has been the subject of theoretical modelling. For instance, \cite{dicintio2019} have found that coplanar and corotating mergers together with aligned accretion of gas at early times could lead to the formation of discs with a small bulge \citep[e.g.,][]{sales2012}. The authors have also found that the spatial extent of star formation (SF) is directly linked to whether a galaxy is high surface brightness (HSB) galaxy or low surface brightness (LSB) galaxy. HSBs (presence of a central bulge) form their new stars in the central parsecs, whereas LSBs (absence of a central bulge) exhibit significantly more extended star formation: stars younger than 5 Gyr are spread throughout the disc.

Supernovae (SNe) are an additional component that may contribute to the formation of BL galaxies. Gas is injected or heated into the interstellar medium through the death of massive stars, thereby regulating the star formation of galaxies. Specifically, recently, BL dwarf disc galaxies have recently been generated through simulations that incorporate SN feedback \citep[e.g.,][]{governato2010,teyssier2013} as well as in disc galaxies \citep{rosasguevara2025}. However, the impact of SN feedback on the formation or absence of bulges in galaxies with the MW mass remains poorly understood.

Cosmological hydrodynamical simulations \citep[e.g.,][]{schaye2015,pillepich2018b} have enabled detailed studies of bulge and disc formation. Some works have focused on how mergers and gas accretion shape these components \citep{wu2025}, while others have explored the formation of BL galaxies \citep{rodriguez2025}. For example, \citet{wu2025} find that certain orbital configurations, such as coplanar mergers, favour disc survival. \citet{rodriguez2025} show that early aligned gas accretion supports the formation of the most rotationally supported discs. Both studies were based on the TNG100 simulation, but now, with the higher resolution granted by TNG50, a more detailed analysis can be performed. Focusing on MW- and M31-like galaxies, \citet{sotillo2022} show that many systems experience major mergers yet retain or rebuild their stellar discs. These galaxies often exhibit thicker discs, more accreted stars, enhanced gas fractions, and active star formation, yet their disc morphologies can survive or reform despite these disruptions. \citet{pinna2024a} study the formation and evolution of discs with AURIGA simulations, finding that the early stages of galaxy evolution, influenced by gas accretion, dictate the initial mass of the disc. Consequently, the sensitivity of the disc to additional processes and the recent merger history emerge as critical factors for disc formation, evolution, and survival.

Nevertheless, previous observational and numerical studies have not provided a unified picture of the diverse evolutionary pathways that can lead to the formation of BL galaxies seen as MW galaxies. In this work, we investigate these pathways using the catalogue of MW and M31 analogues in TNG50 \citep{pillepich2023,sotillo2022}. This catalogue includes galaxies with $M_{\rm halo}<10^{13}\,\Msun$ and $10^{10.5}\Msun< M_*<10^{11.2}\Msun$, which are selected to be isolated and visually disky, or with axial ratios $c/a < 0.45$. We further applied a morphological selection based on a bulge-to-disc mass ratio of $B/D \leq 0.08$, consistent with MW estimates and theoretical predictions for secularly built bulges \citep{shen2010}. For comparison, we also selected a sample of bulge-dominated (BD) galaxies with $B/D>1$.

This study is motivated by the Bulgeless Evolution And the Rise of Discs (BEARD) survey (Mendez-Abreu et al. in prep)\footnote{\url{https://beardsurvey.com}}, an observational programme designed to constrain the formation histories of MW analogues. BEARD targets a volume-limited ($\leq 40$~Mpc) sample of 50 massive BL spiral galaxies ($M_* \geq 10^{10}\,\Msun$) selected using a similar morphological criterion to this work, including $B/T < 0.1$ and $B/D < 0.08$ (Zarattini S., et al. in prep)

This paper is organised as follows. In Section~\ref{sec:method}, we describe the simulations and our selection criteria for BL and BD galaxies. In Section~\ref{sec:results}, we compare their merger histories and mass assembly of the two samples and link them and connect them with their $z=0$ galaxy properties. In Section~\ref{sec:galev}, we investigate the evolution of their dynamical components, the parameters of major mergers, and the age of the stellar populations. We discuss our findings in Section~\ref{sec:discussion} and summarise our conclusions in Section~\ref{sec:summary}.

  \section{Methodology}
  \label{sec:method}
  In this section, we describe the galaxy sample selection in Section~\ref{sub:discsample}. We also describe merger identification in Section~\ref{subsec:mergersparam} based on the TNG50 simulation, which is briefly described in Appendix~\ref{app:TNG50sim}.

  \begin{table}
  \caption{Galaxy samples.}
 \label{table:samples}
  \centering
  \begin{tabular}{cccccc}
  \hline
  Name                                & $n_{\rm g}$ &  $f_{g}$ & $n_{\rm MM}$     \\
                                      &             &          &                                                     \\

  \hline
  \hline
  PS (parent sample)       &  $198$      &   $1.0$  &    $94$            \\
  BL ($B/D\leq0.08$)     &  $31$       &   $0.15$ &    $14$                   \\
  BD ($B/D>1$)     &  $32$       &   $0.16$ &     $16$        \\

  \hline
  \end{tabular}
  \tablefoot{Parent sample (PS, \citealt{pillepich2023}), Bulgeless (BL) and bulge-dominated (BD). BL and BD samples are distinguished based on the stellar mass ratio of the bulge to the thin disc ($B/D$). Columns from left to right list: name of the sample, number of galaxies in each subsample, fraction relative to PS, number of galaxies which experienced at least one major merger since $z=2$.}
  \end{table}

  \begin{figure*}
  \centering

  \begin{subfigure}[t]{0.49\textwidth}
    \centering
    \includegraphics[width=\textwidth]{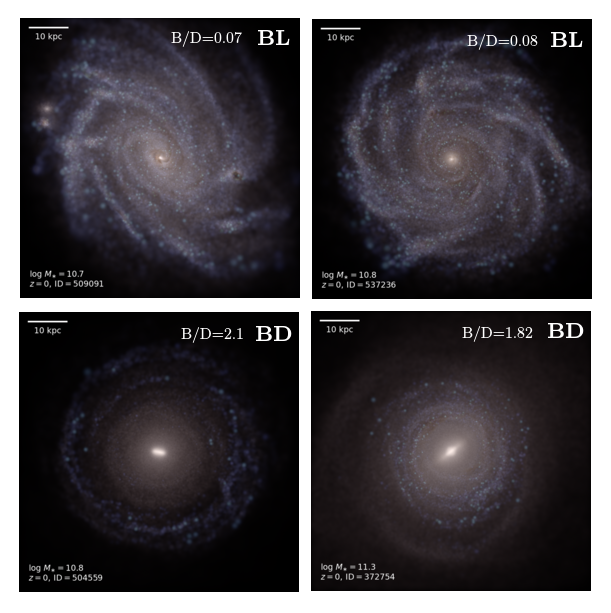}
    \caption{}
    %\caption{}
    \label{fig:mockimagesa}
  \end{subfigure}\hfill
  \begin{subfigure}[t]{0.49\textwidth}
    \centering
    \includegraphics[width=\textwidth]{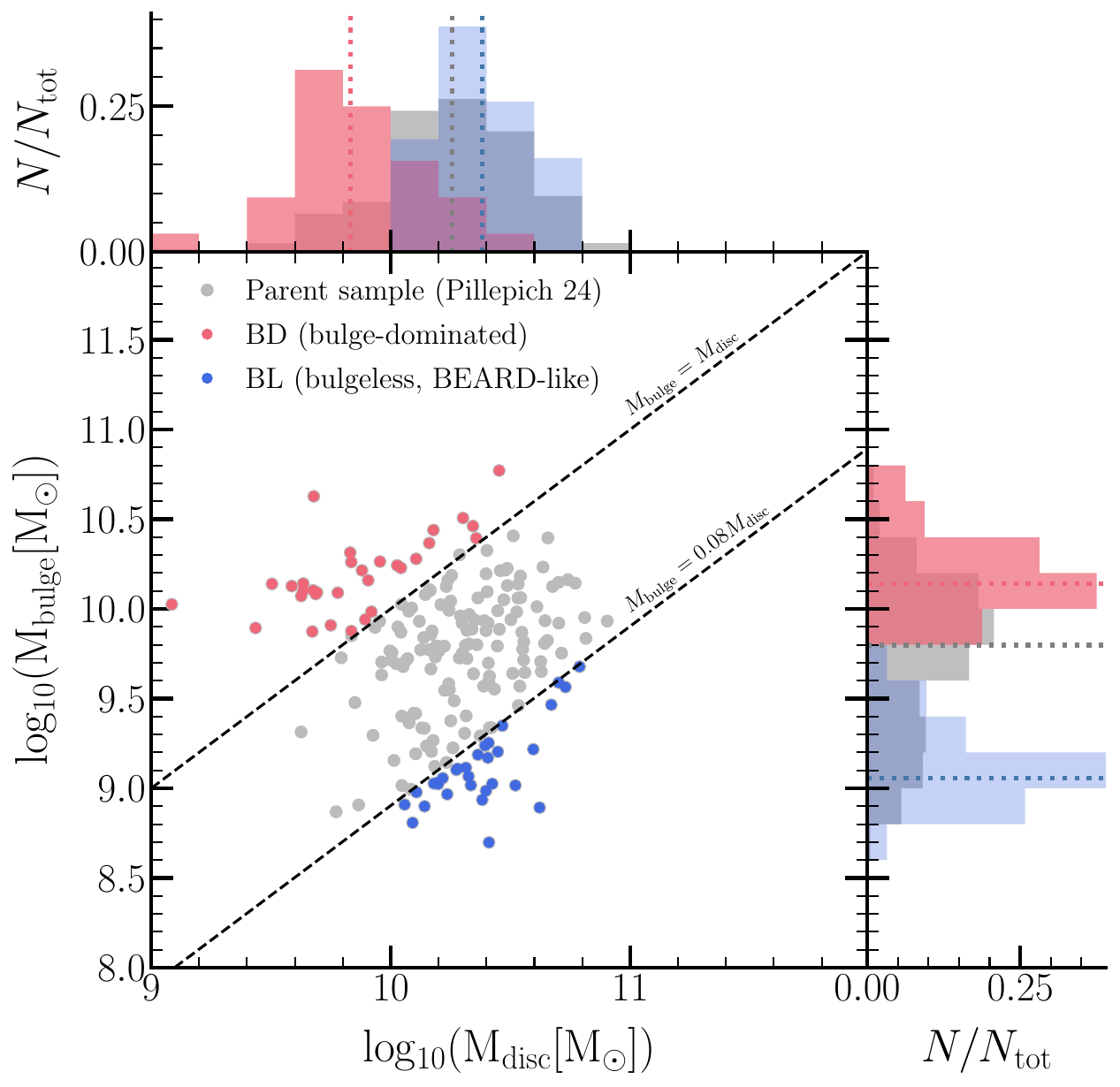}
    \caption{}
    \label{fig:mockimagesb}

  \end{subfigure}
\caption{Examples and stellar masses of galaxy samples.(a) Composite stellar light image of three HST/ACS bands (F435W, F606W, F775W) from \protect\citealt[][(TNG database)]{pillepich2023} for bulgeless (top) and bulge-dominated (bottom) galaxies. Galaxies in the left panels have not experienced a major merger since $z=2$. (b) Bulge mass as a function of thin disc mass for BL (blue), BD (magenta), and  PS (grey) samples.}
  \label{fig:mockimages}
\end{figure*}

\subsection{Bulgeless, bulge-dominated, and parent samples}
\label{sub:discsample}

We focused our analysis on the sample of MW and M31 analogues (hereafter parent sample, PS) in TNG50 introduced by \cite{pillepich2023}.
%\footnote{\url{https://www.tng-project.org/data/milkyway+andromeda/}}.
The sample satisfies the following conditions at $z=0$: (i) the stellar mass, $M_{*}$ (measured within an aperture of $30$ kpc), lies between  $10^{10.5}\Msun$ and $10^{11.2}\Msun$, and the host halo mass lies below < $10^{13}\Msun$; (ii) the stellar disc has a minor-to-major axis ratio of $c/a \leq 0.45$ (measured between $1-2$ half-stellar mass radii) or it appears visually disc-like with obvious spiral arms ($25$ galaxies with $c/a>0.45$ are included); (iii) no other galaxies with  $M_{*} \geq 10^ {10.5}\Msun$ are within a distance of $500$ kpc. The catalogue consists of $198$ galaxies.

Additionally, we use the \textsc{mordor} dynamical decomposition code \citep{zana2022} to identify up to five stellar components, including the bulge, thin disc, thick disc, stellar halo and secular bulge, based on stellar circularity ($\epsilon$) and binding energy ($E$). Further details are provided in Appendix~\ref{ap:mordor}.

We define BL (also BEARD-like)  galaxies as those for which the bulge mass is less than  $8\%$  of the thin disc mass ($B/D\leq0.08$), the expected value for the MW \citep{shen2010}. We focus solely on the thin disc mass to confirm that the disc constitutes the dominant component; however, the disc mass could be greater. The selection results in $31$ galaxies corresponding to $15\%$ of the original PS catalogue. This fraction is very similar to the fraction obtained by \cite{du2020,pillepich2023} ($25$ galaxies, $12\%$) using a different technique to identify galaxy components. We also select galaxies with a bulge mass that is greater than their thin disc mass ($B/D>1$), namely, bulge-dominated (BD)  galaxies. This corresponds to $32$ galaxies ($16$ per cent) of the total PS catalogue. Table \ref{table:samples} presents further details on the three samples, including the number of galaxies that have experienced at least one major merger.

The composite stellar light images of three HST/ACS bands (F435W, F606W, F775W, \citealt{pillepich2023}) for two BL galaxies (top) and two BD galaxies (bottom) are depicted in Fig. \ref{fig:mockimagesa}. The central bulge component of all examples is substantially different, despite the presence of massive discs. Galaxies in the left column do not experience a major merger since $z=2$, whereas the galaxies in the right column do. Fig. \ref{fig:mockimagesb} shows the relation between the bulge and disc masses, highlighting that BL galaxies tend to have the most massive discs and the least massive bulges, in contrast to BD galaxies.

\subsection{Merger parameters}
\label{subsec:mergersparam}

 Galaxies are tracked over time by the \textsc{Sublink} merger tree algorithm \citep{rodriguezgomez2015}.  Galaxies that went through mergers have more than one progenitor, and for our purpose, we track the most massive progenitors of merged galaxies. The merger trees are found in the public database of TNG simulations with 100 epochs, with a temporal interval between successive snapshots varying from  $0.07$ Gyr to $0.15$ Gyr, and we focus on the merger trees with additional conditions in the secondary galaxies described in \cite*{sotillo2022}. We split mergers into major and minor mergers.  Major mergers are distinguished by a stellar mass ratio of the secondary galaxy to the primary galaxy, denoted as $\mu=M_{*}^{s}/M_{*}^{p}$, that exceeds $0.25$, while minor mergers are defined by $\mu$ values that range from $0.01$ to $0.25$. The stellar mass ratio is calculated at the snapshot of the maximum stellar masses of the secondary galaxy. We trace the history of each galaxy and count the number of major and minor mergers experienced.

In addition to distinguishing between major and minor mergers, we estimate the gas mass ratio of the merger, $\mu_{\rm gas}$, with the goal of measuring the gas involved during the merger event, which is defined as:
\begin{equation}
  \mu_{\rm gas} = (M_{\rm gas}^{s}+M_{\rm gas}^{p})/( M_{\rm bary}^{s}+M_{\rm bary}^{p}),
\end{equation}
where $M_{\rm gas}^{s}$ and $M_{\rm gas}^{p}$ correspond to the gaseous masses of the secondary and primary galaxies, respectively, while $M_{\rm bary}^{s}$ and $M_{\rm bary}^{p}$  are the corresponding baryonic masses (gas and stellar components) within a twice stellar half mass radius at the time when the secondary has the maximum stellar mass.

To determine the orientation of the mergers, we calculate the mass-weighted stellar-specific angular momentum as $j_{*} =J_{*}/M_{*}$  where $J_{*}$ is the angular momentum (AM) of the stellar component of each galaxy. We also calculate the orbital AM vector of the secondary galaxy defined as  $\vec{j}_{\rm orbital}= \vec{r} \times \vec{v}$ where the position ($\vec{r}$) and velocity vector ($\vec{v}$) of the secondary galaxy are calculated relative to the rest reference system of the primary galaxy in the last snapshot when the two galaxies are identified as distinct objects. Then, we calculate two angles: (1) $\theta_{\rm spin}$, is the angle subtended between the $\vec{j}_{*}$  vectors of the two galaxies that are at the point of merging, and (2) $\theta_{\rm orb}$, is the angle between $\vec{j}_{*}$  of the primary galaxy and the orbital AM vector ($\vec{j}_{\rm orbital}$). Both angles are defined by:

\begin{equation}
  \theta_{\rm spin} = {\rm acos} \big (\hat{j}_{*}^{s} \cdot \hat{j}_{*}^{p} \big), \\
  \theta_{\rm orb}  = {\rm acos} \big (\hat{j}_{\rm orbital} \cdot \hat{j}_{*}^{p} \big),\\
  \label{eq:angles}
\end{equation}
where  $\hat{j_{*}}^{s}$ and  $\hat{j_{*}}^{p}$ denote the normalised $j_{*}$ vectors of the secondary and primary galaxies, respectively. Both angles allow us to identify the orbital alignment and spin orientation of merging galaxies. In particular, $\theta_{\rm spin}$ quantifies the spin alignment of both galaxies. We define corotating, perpendicular and counterrotating mergers as those with $|\rm cos(\theta_{\rm spin})|>0.7$ (angles between 0-45 degrees), $-0.15<\rm cos(\theta_{\rm spin})<0.15$ (angles between 80-99 degrees), respectively. The angle $\theta_{\rm orb}$ quantifies the orbital orientation of the merging galaxy. Coplanar mergers are those with $|\rm cos(\theta_{\rm orb})|>0.7$ and perpendicular mergers with \textbf{$|\rm cos(\theta_{\rm orb})|\leq 0.3$.}

\section{Linking merger and assembly histories to $z=0$ BL and BD galaxy properties}
\label{sec:results}

We investigate the merger histories and stellar mass assembly of BL, BD, and PS galaxies, in Section \ref{sub:discsample} and Section \ref{subsec:mergers}, respectively. In Section \ref{subsec:galaxylocal}, we investigate the evolution of their properties at $z=0$.

\subsection{Merger histories and stellar mass assembly}
\label{subsec:mergers}

  \begin{figure}

  \includegraphics[width=1\columnwidth]{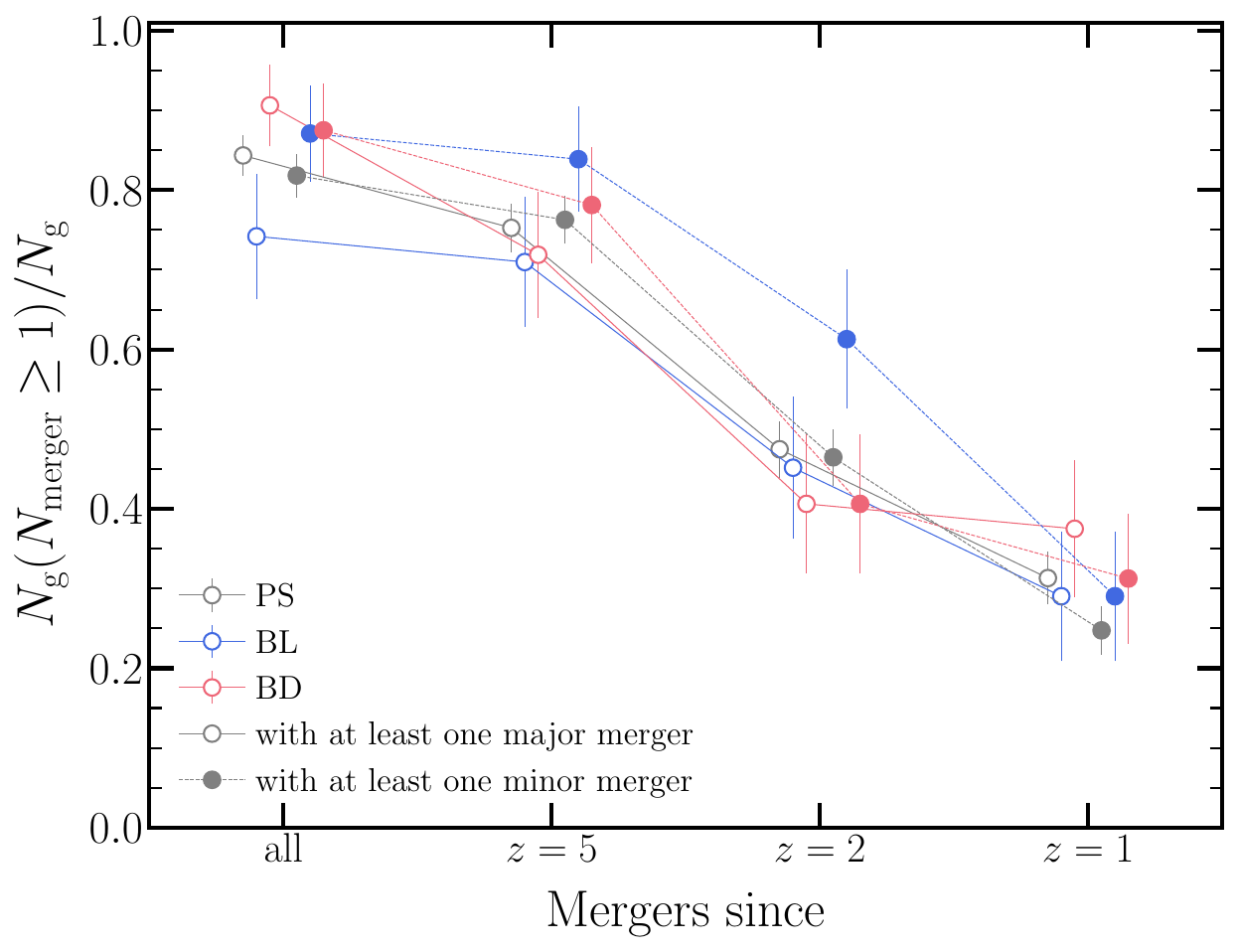}
  \caption{Fraction of galaxies that experienced at least one major (minor) merger since different redshifts, shown for BL, BD, and PS galaxies. Error bars correspond to binomial errors that are calculated using the number of galaxies and those with mergers in each bin as $\sigma =(f_{\rm g}(1-f_{\rm g})/N_{\rm g})^{0.5}$. For clarity, sample lines are slightly offset along the x-axis to avoid overlapping error bars. BL galaxies show the lowest major merger fraction ($0.74$), while BD galaxies exhibit the highest ($0.91$) in their lifetime.}
  \label{fig:mergerhistories}

  \includegraphics[width=1\columnwidth]{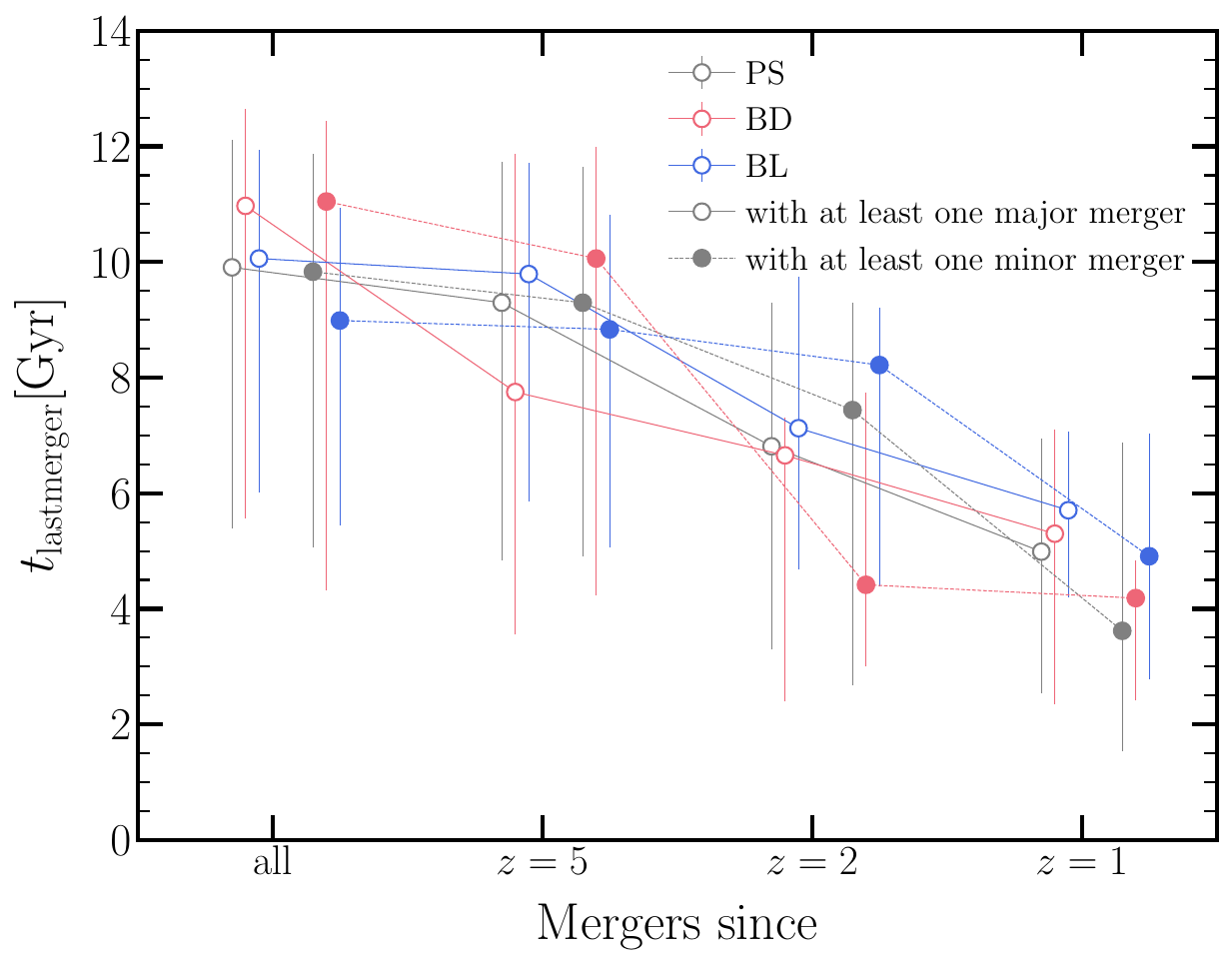}
  \caption{Median lookback time of the last major and minor merger for galaxies that experienced at least one such event, shown for different samples and redshift thresholds.  Error bars indicate the $20^{\rm th}$ and $80^{\rm th}$ percentiles of each distribution. For clarity, sample lines are slightly offset along the x-axis to avoid overlapping error bars.}
\label{fig:lastmerger}
\end{figure}

The fraction of BL, BD, and PS galaxies that experienced at least one major (solid lines with empty circles) or minor merger (dashed lines with filled circles) since different redshifts is shown in Fig.~\ref{fig:mergerhistories}. For all three samples, the fraction increases with redshift, with the lowest values in the recent past ($z<1$) and the highest over their entire lifetime (all). While the overall trend is similar across samples, notable differences appear: BL galaxies show the lowest total fraction of galaxies with at least one major merger ($0.74$), while BD galaxies show the highest ($0.91$), with the parent sample in between ($0.84$). Since $z=2$, at least $45\%$ of BL and $41\%$ of BD galaxies have experienced a major merger (see Appendix~\ref{ap:mh}).

The fraction of galaxies that experienced at least one minor merger in their entire lifetime (label `all' in Fig.~\ref{fig:mergerhistories}) is similar for all the samples (filled circles with dashed lines).This cumulative similarity, however, does not preclude differences in the timing of the minor mergers. For instance, BL galaxies have, on average, a higher fraction of systems that experience at least one minor merger after $z=2$ than their bulge-dominated counterparts, although the scatter is high and the difference becomes smaller at $z=5$.

The median lookback times of the last major and minor mergers for each sample, considering only galaxies that experienced at least one major or minor merger, are shown in Fig.~\ref{fig:lastmerger}. The median lookback time for the last major merger is similar across samples, except for BD galaxies, which have slightly earlier mergers ($11.0$ Gyr ago) compared to BL galaxies ($10.1$ Gyr ago) when considering all mergers in their lifetime. We note, however, that the distributions show substantial scatter.

Conversely, we observe more significant differences in the lookback time of minor mergers among the samples, particularly for mergers at $z<2$. The last minor mergers in BD galaxies occurred earlier (around $11.0$ and $10.1$ Gyr ago) compared to BL galaxies (approximately $9.0$ and $8.8$ Gyr ago) when considering all minor mergers and those at $z<5$. However, this trend reverses when focusing on mergers at $z<2$: the median lookback time of the last minor merger in BL galaxies is $8.2$ Gyr, whereas it is more recent, $4.4$ Gyr ago, in BD galaxies.

  \begin{figure}
  \begin{tabular}{c}
  \includegraphics[width=1\columnwidth]{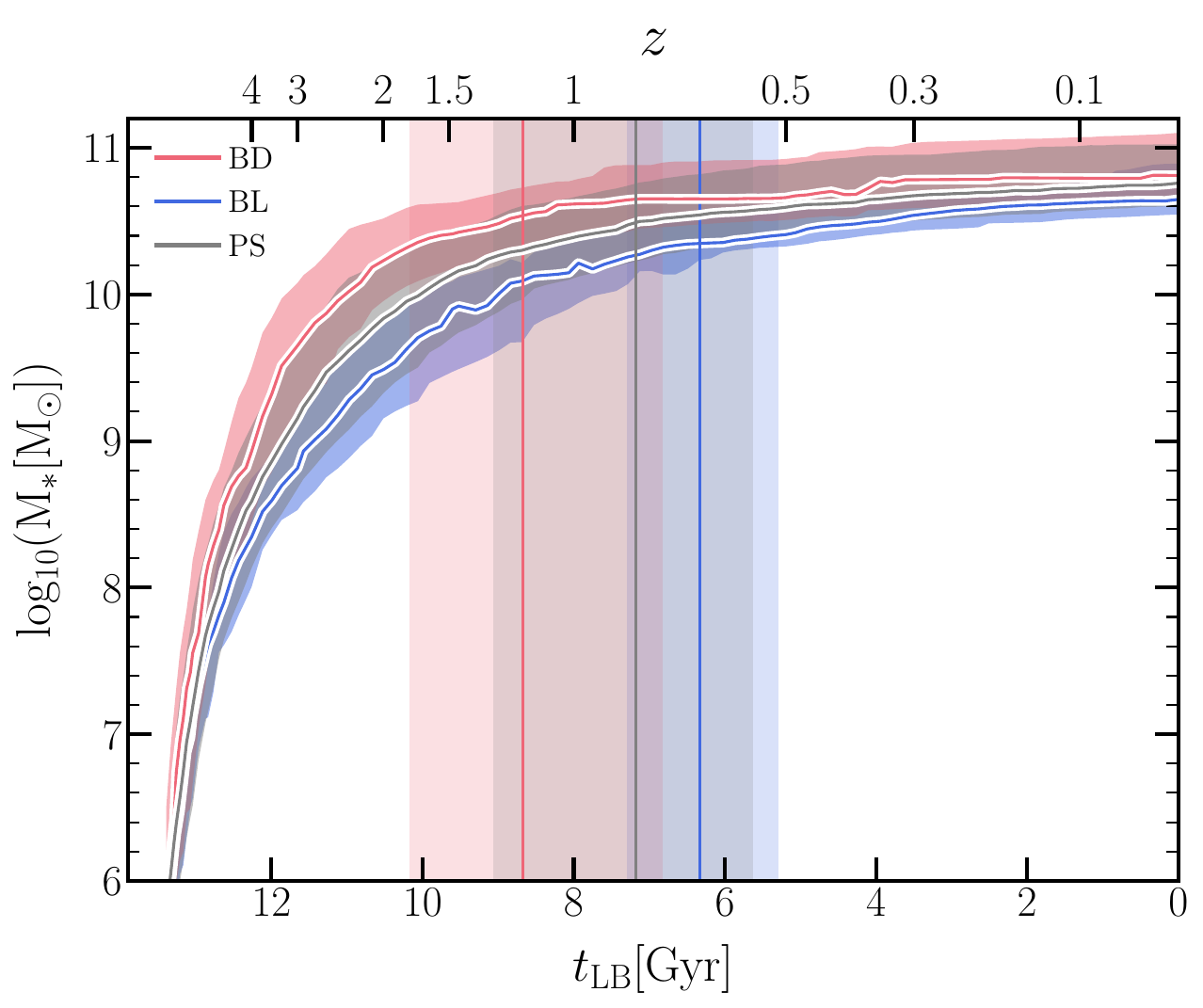} \\
  \end{tabular}
  \caption{Mass assembly of BL, BD, and PS galaxies. Vertical solid lines indicate the redshift (lookback time) when galaxies assembled $50\%$ of their final stellar mass. Coloured regions show the $20^{\rm th}$ and $80^{\rm th}$ percentile range, and solid lines represent the median values. Bulgeless galaxies assemble their mass later than bulge-dominated galaxies.}
  \label{fig:massevolution}
  \end{figure}

Our findings align with those in \cite*{sotillo2022}, which investigate the merger histories of the TNG50 parent sample. Their findings indicate that $85\%$ of the MW analogues have undergone at least one major merger in their lifetime (see their Fig.~2). This agrees with the $84\%$ identified for the PS catalogue.  Our results are also consistent with semi-analytic models of galaxy formation, where BL  galaxies primarily grow by minor mergers and gas accretion \citep[e.g.,][]{izquierdo2019}.

Hereafter, we investigate the galaxies that have undergone major mergers since $z=2$ in order to assess their impact on the evolution of the samples. As illustrated in Fig.~\ref{fig:lastmerger}, the median lookback time of the most recent major merger is comparable in both BL  ($7.1$ Gyr ago) and BD ($6.6$ Gyr ago) galaxies. Moreover, the distribution of the lookback times of the major mergers in both samples is sufficiently broad with $20^{\rm th}$ and $80^{\rm th}$ percentiles ranging from $4.7$ Gyrs to $9.7$ Gyrs ago for BL  galaxies, thereby encompassing estimates for the lookback time of the most recent major merger that occurred in the Milky Way \citep{belokurov2018,helmi2018,naidu2021}. Finally to make sense of the merger statistics in our sample, Table~\ref{table:samples} (see Appendix \ref{ap:mh}, Fig.~\ref{fig:mergerhistoryz2}) shows the number of galaxies that have at least one major merger since $z=2$ for the different samples, in particular we find that $14$ BL  galaxies and $16$ BD galaxies have experienced at least one merger.

The median stellar mass evolution for the BL, BD, and PS samples is shown in Fig.~\ref{fig:massevolution}. All samples grow by roughly two orders of magnitude from $z=4$ to $z=0$, indicating significant stellar mass buildup. However, differences in their assembly histories appear as early as $z=4$, despite similar final stellar masses within the scatter. We define the formation time ($t_{50}$) as the lookback time when a galaxy assembles $50\%$ of its $z=0$ stellar mass. On average, BL galaxies form later, with a median $t_{50} = 6.3$ Gyr ($z_{50} = 0.7$), compared to BD galaxies with $t_{50} = 8.7$ Gyr ($z_{50} = 1.2$) which results in a difference of $2.4$ Gyr. The PS sample shows intermediate values ($t_{50} = 7.2$ Gyr, $z_{50} = 0.8$). BD galaxies also display larger scatter, indicating more diverse assembly histories.

These results suggest that overall BD galaxies build their stellar mass earlier and more rapidly than BL galaxies, which grow more gradually. This is in agreement with previous numerical \citep{pinna2024a} and observational \citep{sattler2023,sattler2025} studies from the star formation histories of discs in MW-mass galaxies of various morphological types and SFRs, where later-type galaxies have had a slower evolution, and this leads to younger average ages and lower metallicities.

\subsection{BL and BD galaxy properties at $z=0$}
\label{subsec:galaxylocal}
  \begin{figure}
  \includegraphics[width=\columnwidth]{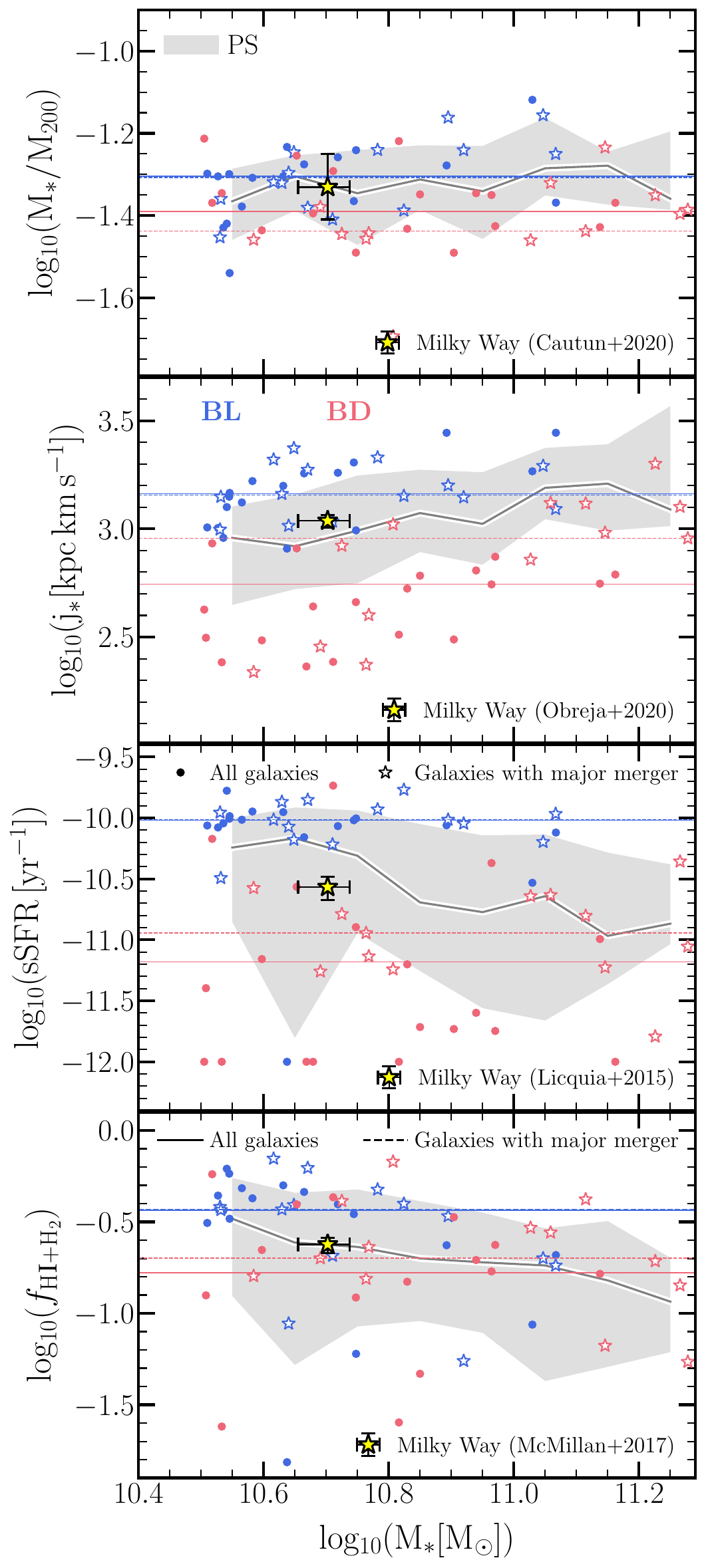}
  \caption{Stellar-to-halo mass ratio, specific angular momentum, sSFR, and H\textsc{i}+H$_2$ fraction as functions of stellar mass. Grey regions and grey solid lines indicate the $20^{\rm th}$-$80^{\rm th}$  percentile range and median values for the PS sample, respectively. Horizontal lines correspond to median values as the legend indicates. For reference,  Milky-Way estimates are included, based on analytic models calibrated to observations from \textit{Gaia} DR2, VERA, and APOGEE. BL galaxies typically reside in lower-mass dark matter haloes, exhibit higher specific angular momentum, H\textsc{i}+H$_2$ gas fractions, and star formation rates than their BD counterparts.}
  \label{fig:galaxyproperties}
  \end{figure}

Several key trends as a function of stellar mass for the three samples are shown in Fig.~\ref{fig:galaxyproperties}. In the top panel, we present the stellar-to-halo mass ratio ($M_*/M_{200}$). BL galaxies are likely to reside in less massive dark matter haloes than BD galaxies, making them, on average, more baryon-dominated systems.
We do not find a clear systematic trend in the stellar-to-halo mass ratio of BL galaxies by considering those that experience at least a major merger, and the median stellar-to-halo mass ratios (blue horizontal solid and dashed lines) remain comparable. In contrast, BD galaxies that experience a major merger tend to exhibit the lowest stellar-to-halo mass ratios across the samples.
Overall, BD galaxies exhibit the lowest stellar-to-halo mass ratios in the sample, even compared to both BL and PS galaxies. Since this ratio serves as a proxy for star formation efficiency, this suggests that haloes hosting BL galaxies are more efficient at converting baryons into stars than those hosting BD systems at fixed halo mass. Supporting this, \cite*{correa2020} found that disc galaxies show higher stellar-to-halo mass ratios than spheroid-dominated systems in hydrodynamical simulations. Similarly, \cite*{scholz2024} reported that halo mass can influence the entire baryonic cycle,  including star formation history, morphology, kinematics, and stellar populations.
For reference, we include the Milky Way stellar-to-halo mass ratio estimate from \citet{cautun2020}, who fit the Gaia DR2 rotation curve using models that incorporate dark matter halo contraction due to baryonic effects. Their analysis yields a value of $\rm log_{10}(M_{*}/M_{200})_{MW}=-1.33$,  which lies within the distribution for both BL  and PS galaxies. Specifically, we find median values of  $\rm log_{10}(M_{*}/M_{200})_{\rm BL }=-1.30$ and $\rm log_{10}(M_{*}/M_{200})_{\rm PS}=-1.30$, while BD galaxies show slightly lower median values of $\rm log_{10}(M_{*}/M_{200})_{\rm BD}=-1.39$ ). All median values lie within the error bars of the Milky Way estimates, showing that all the samples are consistent with the MW stellar-to-halo mass ratio.

One of the key baryonic properties that reflects the dynamical state and evolutionary history of galaxies is the specific stellar angular momentum \citep{du2024}. The second panel of Fig.~\ref{fig:galaxyproperties} shows the total specific stellar angular momentum, $j_{*}$, as a function of stellar mass. As expected for rotationally supported systems, $j_{*}$ increases with stellar mass across all three samples: BL, BD, and PS galaxies. At fixed stellar mass, BL  galaxies exhibit systematically higher specific stellar angular momentum than their BD counterparts. This trend is a direct consequence of our morphological selection criteria. Specifically, BL  galaxies possess more prominent thin disc components ($B/D\leq0.08$) and are therefore more rotationally supported, while BD galaxies ($B/D\geq 1$) exhibit more dispersion-dominated kinematics.

When we examine galaxies with and without mergers, we find no significant difference in $j_{*}$ between BL  galaxies with and without major mergers. In contrast, BD galaxies that have experienced at least one major merger exhibit a higher median specific angular momentum (see the horizontal dotted line in the histogram) than BD galaxies without mergers, though this appears to be driven primarily by higher stellar masses in BD galaxies.
The median values further illustrate these trends: BL  galaxies have a median  $\rm log_{10}(j_{*,\rm BL })=3.16 \,\rm kpc\,\rm km \,s^{-1}$, while BD galaxies show a lower median value of  $\rm log_{10}(j_{*,\rm BD})=2.74 \,\rm kpc\,\rm km\, s^{-1}$. For comparison, we include an estimate of the Milky Way total disc angular momentum from \citet{obreja2022}, who used stellar kinematics data from \textit{Gaia} DR2 and APOGEE to derive the velocity profiles of the thin and thick discs. These were combined with the mass model from \citet{cautun2020} to estimate the halo angular momentum. Stellar populations were separated via clustering in kinematic and chemical abundance space. From  \citet{obreja2022} estimates, we found that $\rm log_{10}(j_{*,\rm MW})=3.03 \,\rm kpc\,\rm km\, s^{-1}$ which is consistent with the median value for PS galaxies ($\rm log_{10}(j_{*,\rm PS})=3.02 \,\rm kpc\,\rm km \,s^{-1}$) and the median for BL galaxies if we consider the mass dependence of $j_{*}$.

The third panel of Fig.~\ref{fig:galaxyproperties} shows the total specific star formation rate ($\rm sSFR=SFR/M_{*}$) as a function of stellar mass for the three galaxy samples. In general, we find that the sSFR decreases with increasing stellar mass for the PS sample, while no clear trend is observed for BD galaxies and a large spread. The relation appears roughly flat with increasing stellar mass for BL  galaxies. Nevertheless, at a fixed stellar mass, BL  galaxies exhibit higher sSFRs than BD galaxies. In fact, a significant fraction of BD galaxies have $\rm sSFR<10^{-11.5}\,\rm yr^{-1}$, categorising them as quenched. We note that for galaxies with $\rm sSFR<10^{-12}\,\rm yr^{-1}$, we set their sSFR to $10^{-12}\,\rm yr^{-1}$ to illustrate the number of quenched systems. When comparing galaxies with and without major mergers, we do not observe a distinct trend as a function of stellar mass. BL  galaxies show similar sSFRs regardless of merger history, while BD galaxies that experienced at least one major merger tend to have higher specific star formation rates, although the increase is less than $0.5$ dex. We also compare these values with the Milky Way estimate presented by \cite{licquia2015}, who used a hierarchical Bayesian statistical framework to derive global properties of the Milky Way, including the star formation rate and stellar mass. By combining various independent measurements from the literature, such as star counts, infrared surveys, and dynamical modelling, and accounting for systematic uncertainties, the authors estimated $\rm sSFR_{\rm MW}=2.71\pm0.59\times10^{-11}\, \rm yr^{-1}$. This value lies within the scatter of the PS sample but is below its median ($\rm sSFR_{\rm PS}=3.31\times10^{-11}\,\rm yr^{-1}$), lower than the median for BL  galaxies ($\rm sSFR_{\rm BL }=9.77\times10^{-11}\,\rm yr^{-1}$), and slightly higher than the median of BD galaxies ($\rm sSFR_{\rm BD}= 6.61\times10^{-12}\,\rm yr^{-1}$). Compared to previous works, \cite{rodriguez2025}, using a broader sample of the most massive rotating galaxies in the TNG100 simulation, find that these galaxies exhibit higher sSFRs relative to those with intermediate morphologies.

Finally, in the bottom panel of Fig.~\ref{fig:galaxyproperties}, we present the gas fraction of hydrogen as a function of stellar mass. We define the gas fraction as $f_{\rm H\textsc{i}+H_{2}} = M_{\rm H\textsc{i}+H_{2}} / M_{*}$, where $M_{\rm H\textsc{i}+H_{2}}$ is the combined mass of atomic and molecular hydrogen. The masses in both atomic and molecular hydrogen were computed for individual gas cells in the TNG simulations by \citet{diemer2019}. This work is based on the method developed by \citet{lagos2015} to estimate neutral gas column densities. The molecular hydrogen mass ($\rm H_{2}$) is calculated following the prescription of \citet{krumholz2013}, in which the transition from atomic to molecular hydrogen depends on the total neutral gas column density, gas-phase metallicity, and the intensity of the interstellar radiation field. As shown in Fig.~\ref{fig:galaxyproperties}, on average, BL  galaxies exhibit higher gas fractions ($\log_{10}(f_{\rm H\textsc{i}+H_{2}}) = -0.43$) than BD galaxies ($\log_{10}(f_{\rm H\textsc{i}+H_{2}}) = -0.78$). This is consistent with BLs having higher star formation efficiency and higher star formation activity as observed previously.  When considering only BL galaxies that experienced a major merger, no significant difference in the median gas fraction is observed. In contrast, BD galaxies that have undergone a major merger tend to have higher gas fractions, with values approaching the median of the PS sample ($\log_{10}(f_{\rm H\textsc{i}+H_{2}}) = -0.67$). For comparison, we include the estimates for the MW from \citet{mcmillan2017}, who model the galactic mass distribution with a detailed accounting of gas components. The MW hydrogen gas fraction lies within the range observed for the PS sample.
 \begin{figure*}
  \centering
  \begin{subfigure}[b]{0.48\textwidth}
    \includegraphics[width=\textwidth]{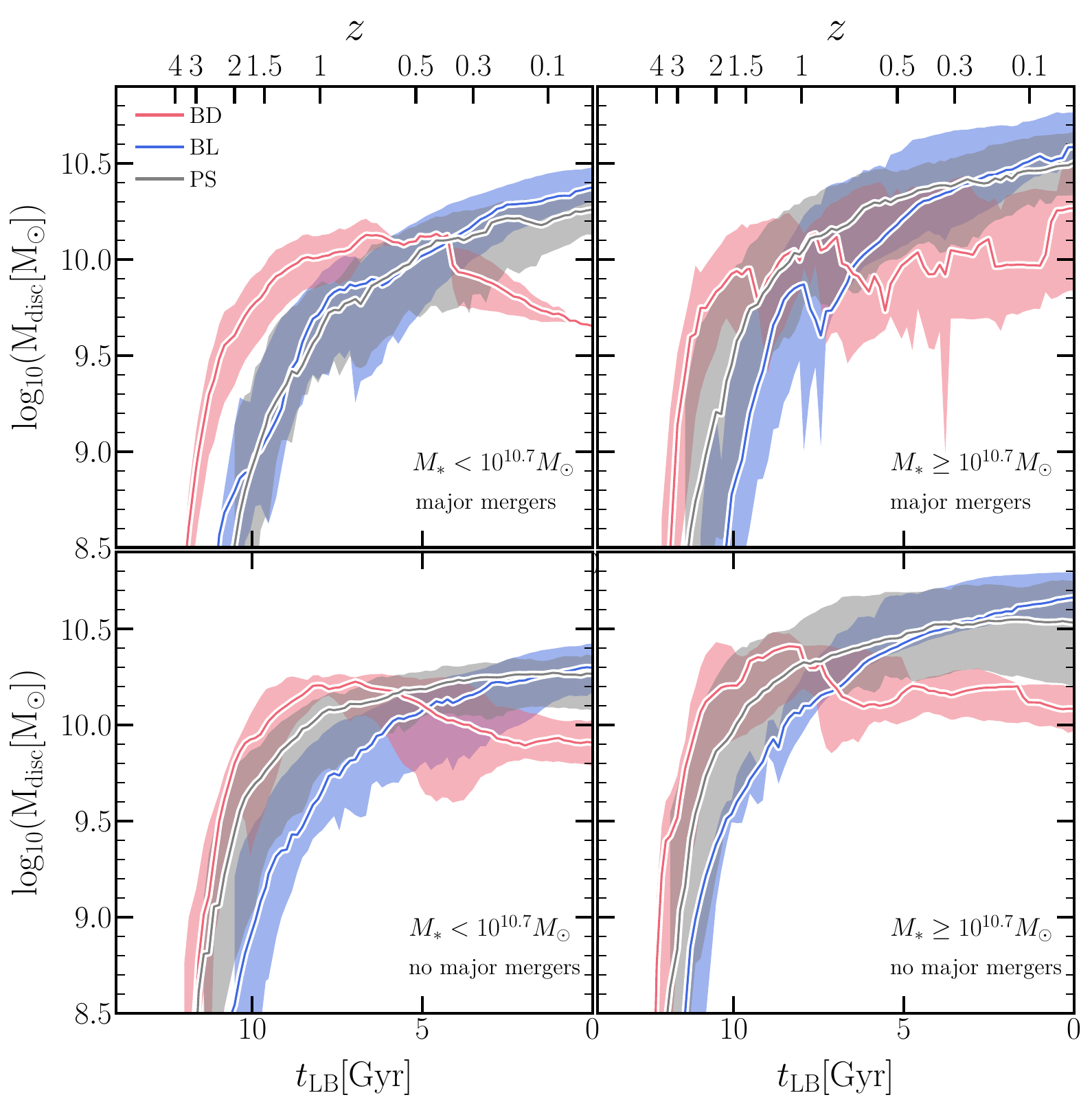}
    \caption{Thin disc mass evolution.}
    \label{fig:evolutionalla}
  \end{subfigure}
  \hfill
  \begin{subfigure}[b]{0.48\textwidth}
    \includegraphics[width=\textwidth]{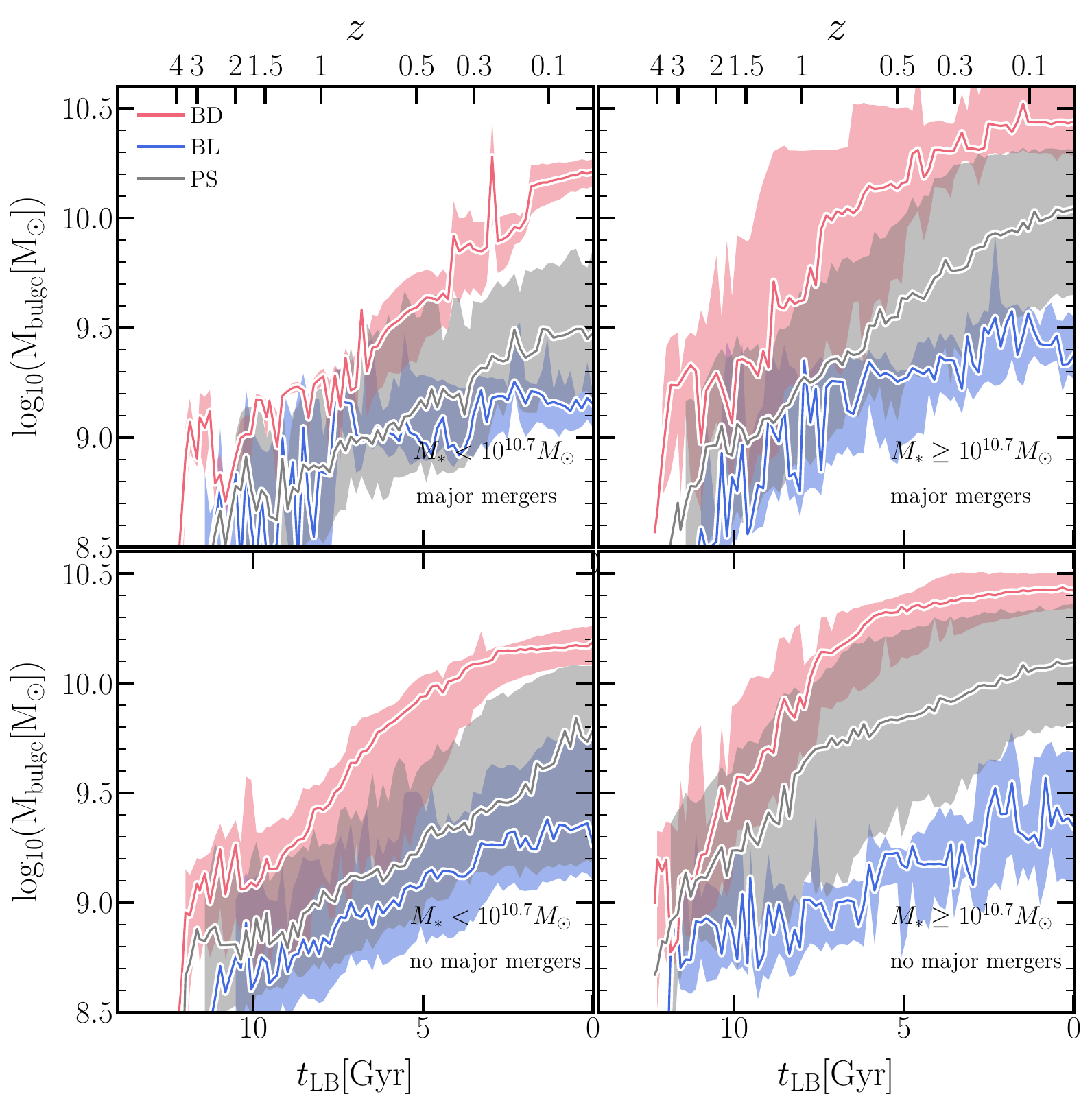}
    \caption{Bulge mass evolution.}
    \label{fig:evolutionallb}
  \end{subfigure}
  \vskip\baselineskip
  \begin{subfigure}[b]{0.48\textwidth}
    \includegraphics[width=\textwidth]{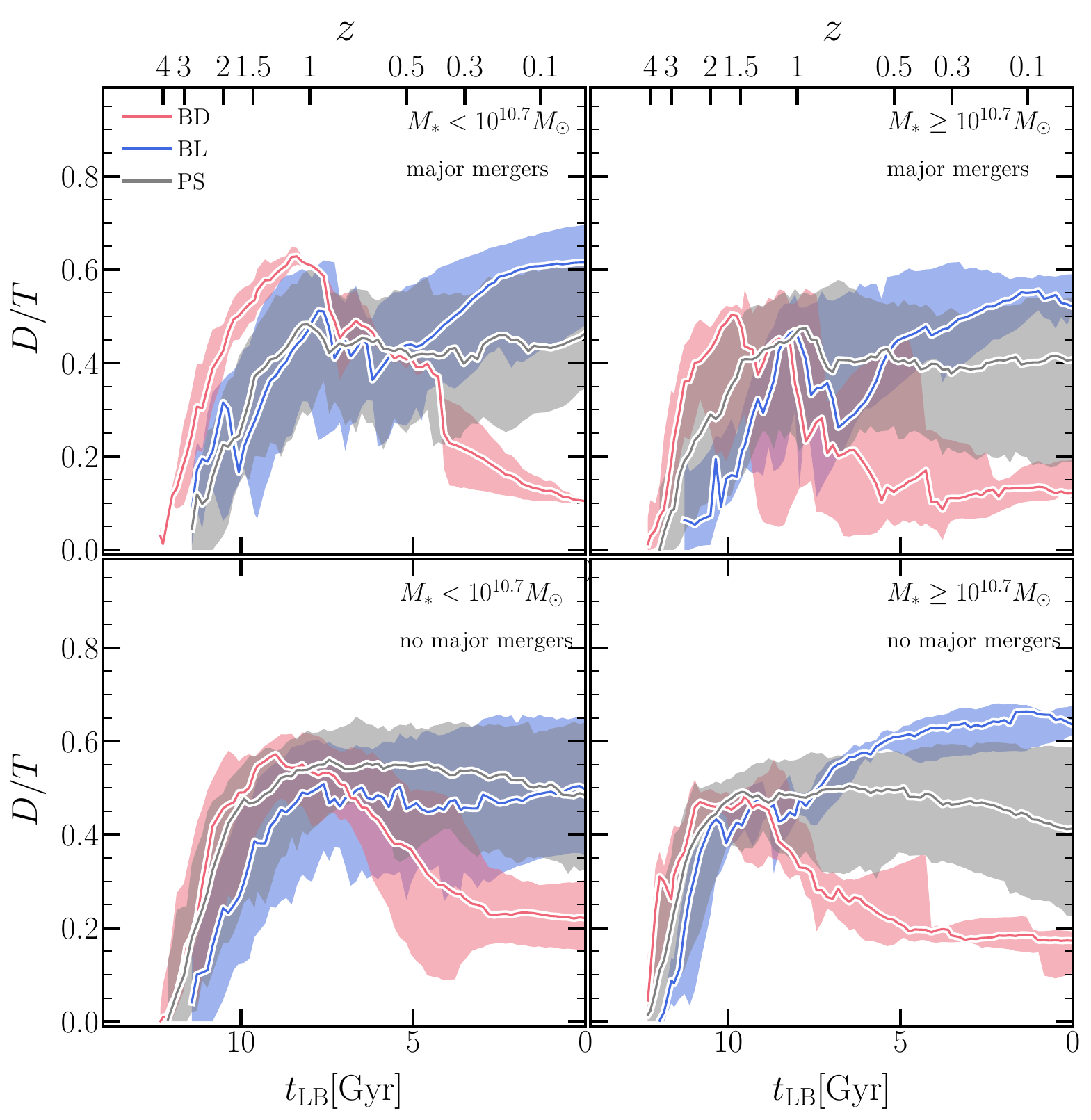}
    \caption{ Thin disc-to-total mass fraction evolution.}
    \label{fig:evolutionallc}
  \end{subfigure}
  \hfill
  \begin{subfigure}[b]{0.48\textwidth}
    \includegraphics[width=\textwidth]{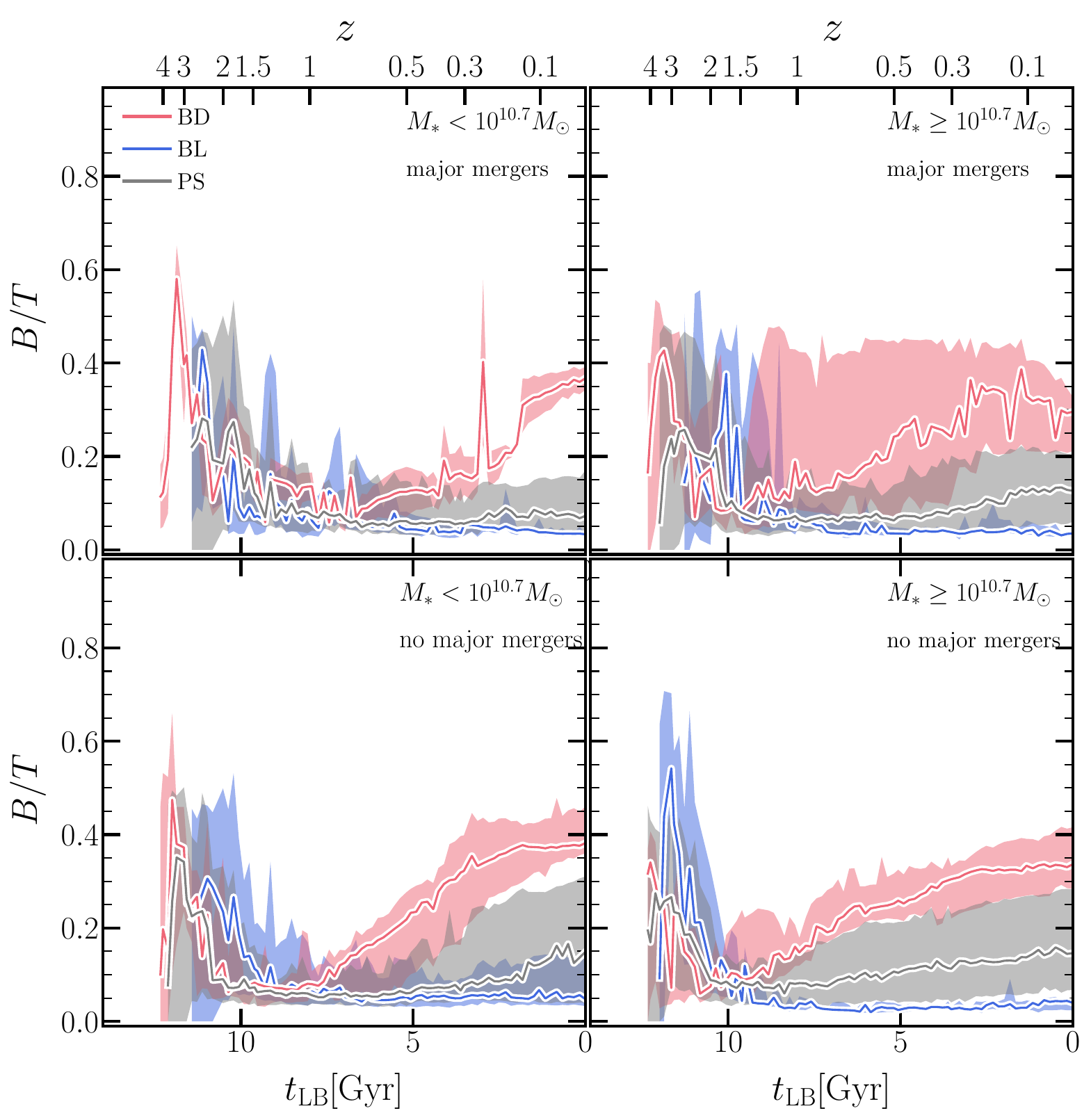}
    \caption{Bulge-to-total mass fraction evolution.}
    \label{fig:evolutionalld}
  \end{subfigure}
  \caption{Evolution of bulge and disc masses and their corresponding mass fractions for the three galaxy samples, shown in bins of total galaxy stellar mass and separated by whether the galaxies experienced at least one major merger at $z<2$. Each bin contains at least six galaxies, except for BD galaxies with major mergers in the low stellar mass bin, which includes only two. Solid lines represent the median values, while shaded regions indicate the $20^{\rm th}$--$80^{\rm th}$ percentile range. BL galaxies tend to build their thin discs at later times compared to BD galaxies, which exhibit a decline in disc mass over time. In high-mass BL systems, disc structures may be temporarily disrupted by mergers but typically recover at later stages.}
  \label{fig:evolutionall}
\end{figure*}

\section{Galaxy evolution: Dynamical components, mergers, and age of the stellar populations}
\label{sec:galev}
In this section, we study the redshift evolution of the bulge and thin disc masses in subsection \ref{subsec:bdevolution}, the merger parameters in section \ref{subsec:mergersplay} and the age distributions of the stellar populations in section \ref{subsec:sfh}.

\subsection{Evolution of galaxy dynamical components}
\label{subsec:bdevolution}
We now examine the redshift evolution of the thin disc and bulge components across the three galaxy samples that were described in Subsection \ref{sub:discsample}. The evolution of thin disc and bulge masses is shown in Fig.~\ref{fig:evolutionalla} and \ref{fig:evolutionallb}, where galaxies are split into two stellar mass bins and further divided based on whether they have experienced at least one major merger at $z<2$.

Across all panels, regardless of stellar mass or merger history, BL  galaxies exhibit a gradual increase in thin disc mass, with approximately $50\%$ of the final disc mass in place by $z\sim0.5$. As shown in Fig.~\ref{fig:evolutionallc}, these systems typically become disc-dominated over time, reaching $D/T\approx0.5$. An exception occurs in high- and low-mass BL  galaxies that undergo a major merger, where the disc may be temporarily disrupted but subsequently recovers after the major merger. It should be noted that the thin disc component is used here, as it represents a lower mass limit of the disc. As illustrated in Appendix~\ref{app:morphev}, the BL galaxies achieve  $(D/T)_{\rm tot}\gsim0.7$ values when all rotational components are considered.

In contrast, BD galaxies show rapid disc growth at earlier times, with their thin discs forming by $z\approx2$ on average and reaching a median stellar mass above $10^{10}\,\Msun$ by $z\sim0.8$. In the low-mass regime, BD galaxies often become disc-dominated earlier ($D/T>0.5$, see Fig.~\ref{fig:evolutionallc}). However, in the most massive BD galaxies that have experienced at least one major merger, disc growth is more irregular, likely due to merger-driven perturbations which heat the disc and increase the mass of other components such as the halo and bulge mass (see Appendix~\ref{app:timeMM}). These galaxies generally recover their disc mass at later times, though in some cases, the disc mass declines by up to $0.5$ dex. At $z = 0$, the median disc contribution reaches $D/T\sim0.10$ for systems with major mergers.
It is evident that the thin disc in BD galaxies without mergers also exhibits a decrease in mass with $D/T\sim0.15$ at $z=0$. It is important to note that these galaxies formed earlier than the other samples, forming a massive disc component at an earlier stage (see Figs.~\ref{fig:evolutionalla} and \ref{fig:evolutionallc}). This, combined with minor mergers occurring later, could easily disrupt the massive disc. In fact, a higher fraction (0.42) of BD galaxies without major mergers undergo at least one minor merger later in time compared to BD galaxies with major mergers that have experienced at least one minor merger (0.38).

The bulge component in BL galaxies shows modest growth, primarily occurring at early times, and reaches masses just above $10^9\,\Msun$, as expected. In contrast, the bulges of BD galaxies grow more rapidly, particularly after $z\sim1$, reaching masses exceeding $10^{10}\,\Msun$. Among low-mass BD galaxies, we do not observe  significant differences in the final bulge mass between systems that have and have not experienced a major merger (see left-column panels of Fig.~\ref{fig:evolutionallb}).

Interestingly, the divergence in bulge mass between BL and BD galaxies appears as early as $z\sim4$, although their bulge mass fractions remain comparable until around $z\sim1.0$. After this point, the bulge-to-total ratio ($B/T$) in BD galaxies increases to values between $0.35$ and $0.45$ (see Fig.~\ref{fig:evolutionalld}), indicating substantial bulge growth relative to the disc. As a result, BD galaxies exhibit $B/D>1$ at low redshift, while BL galaxies show minimal bulge growth since $z\sim1.5$, maintaining $B/T<0.1$ and ending with $B/D\leq0.08$ at $z=0$, due to ongoing disc growth. For comparison, we also include the evolution of the disc and bulge components in the parent sample (PS), which follows an intermediate evolutionary path between the BL and BD populations.
To assess the role of major mergers in shaping these evolutionary pathways, we examine the merger parameters in the following section.

\subsection{How mergers come into play}
\label{subsec:mergersplay}

  \begin{figure}
  \centering
  \begin{subfigure}[b]{0.46\columnwidth}
    \includegraphics[width=\linewidth]{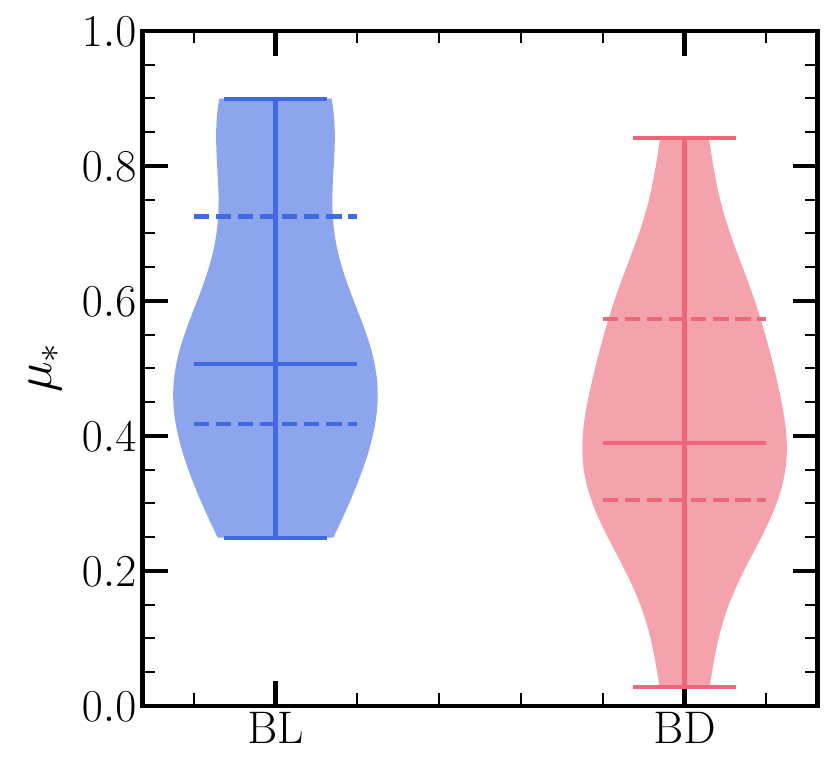}
    \caption{Stellar mass ratio}\label{fig:mergersparam:a}
  \end{subfigure}
  \hfill
  \begin{subfigure}[b]{0.46\columnwidth}
    \includegraphics[width=\linewidth]{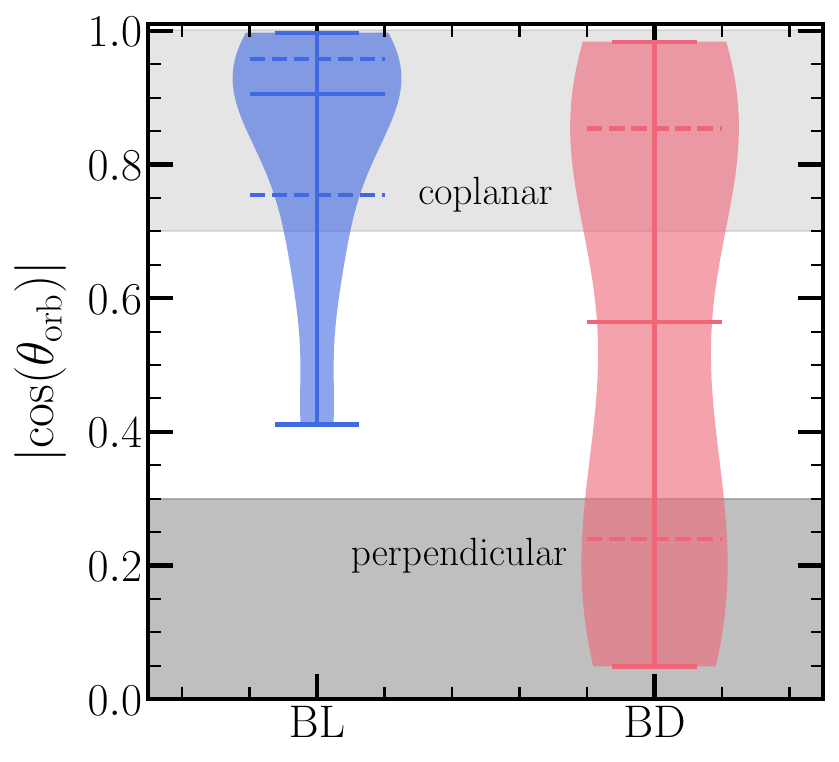}
    \caption{$|\cos\theta_{\rm orb}|$}\label{fig:mergersparam:b}
  \end{subfigure}
  \vskip\baselineskip
  \begin{subfigure}[b]{0.46\columnwidth}
    \includegraphics[width=\linewidth]{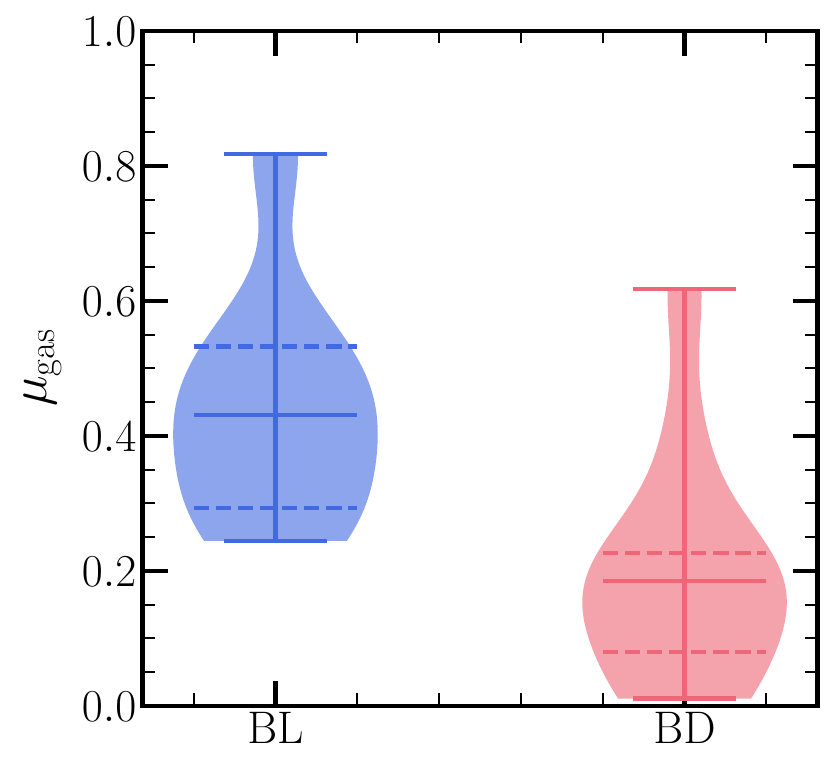}
    \caption{Gas mass fraction}\label{fig:mergersparam:c}
  \end{subfigure}
  \hfill
  \begin{subfigure}[b]{0.46\columnwidth}
    \includegraphics[width=\linewidth]{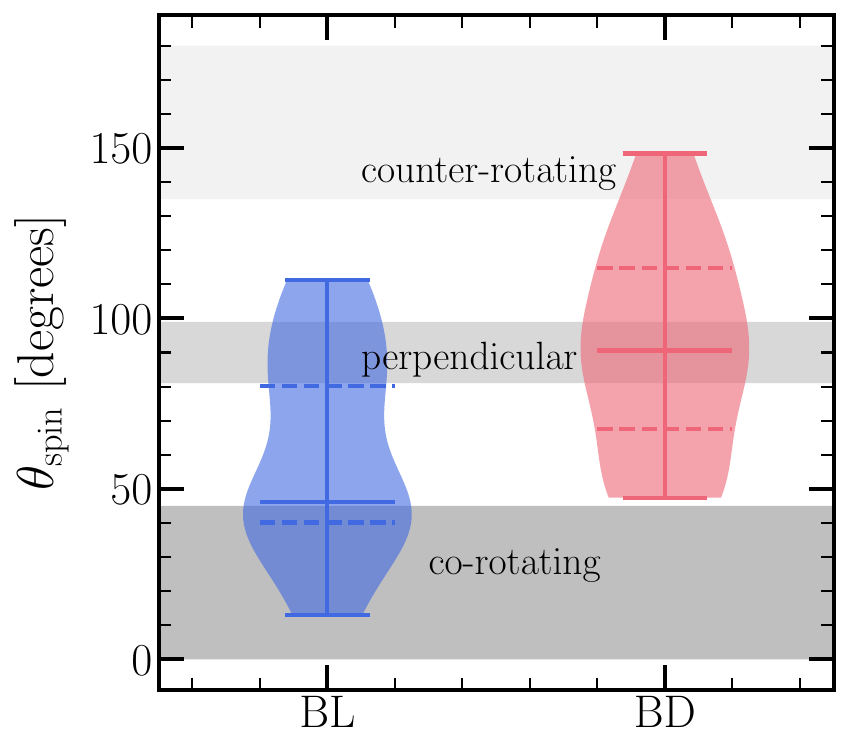}
    \caption{$\cos\theta_{\rm spin}$}\label{fig:mergersparam:d}
  \end{subfigure}

  \caption{
    Probability distribution functions of merger properties for BL and BD galaxies. Panels~\subref{fig:mergersparam:a}–\subref{fig:mergersparam:d} show the distributions of stellar mass ratio, $|\cos\theta_{\rm orb}|$, gas mass fraction, and $\cos\theta_{\rm spin}$, respectively, for major mergers (as defined in Section~\ref{subsec:mergersparam}). Solid and dashed horizontal lines indicate medians and interquartile ranges. Grey hatched regions mark classification thresholds: coplanar mergers have $|\cos(\theta_{\rm orb})| > 0.7$; perpendicular ones have $|\cos(\theta_{\rm orb})| \leq 0.3$; corotating mergers satisfy $\cos(\theta_{\rm spin}) > 0.7$, perpendicular mergers fall in $-0.15 < \cos(\theta_{\rm spin}) < 0.15$, and counterrotating mergers have $\cos(\theta_{\rm spin}) < -0.7$.}
  \label{fig:mergersparam}
\end{figure}

  \begin{figure}
  \begin{tabular}{c}
  \includegraphics[width=1\columnwidth]{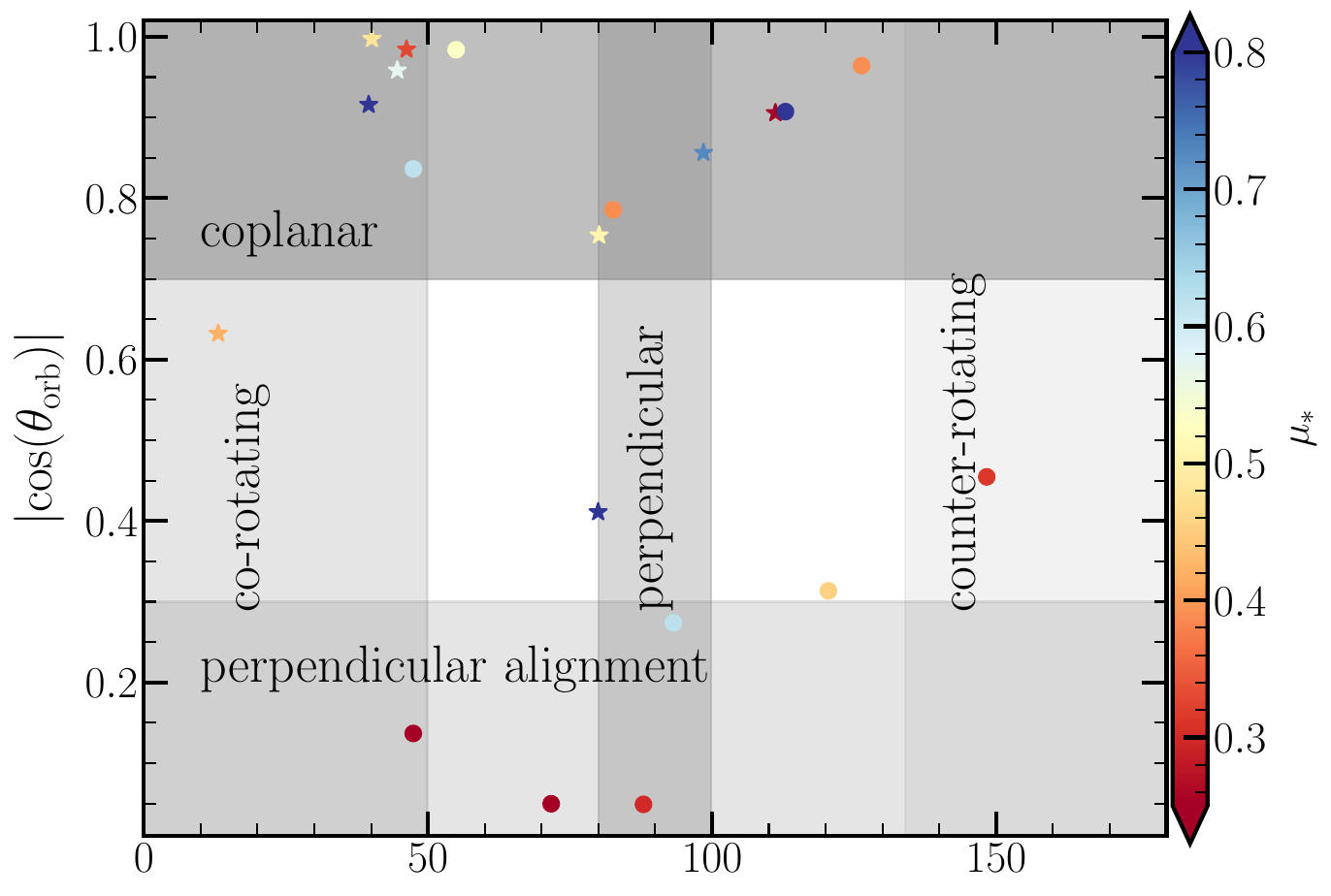} \\
  \includegraphics[width=1\columnwidth]{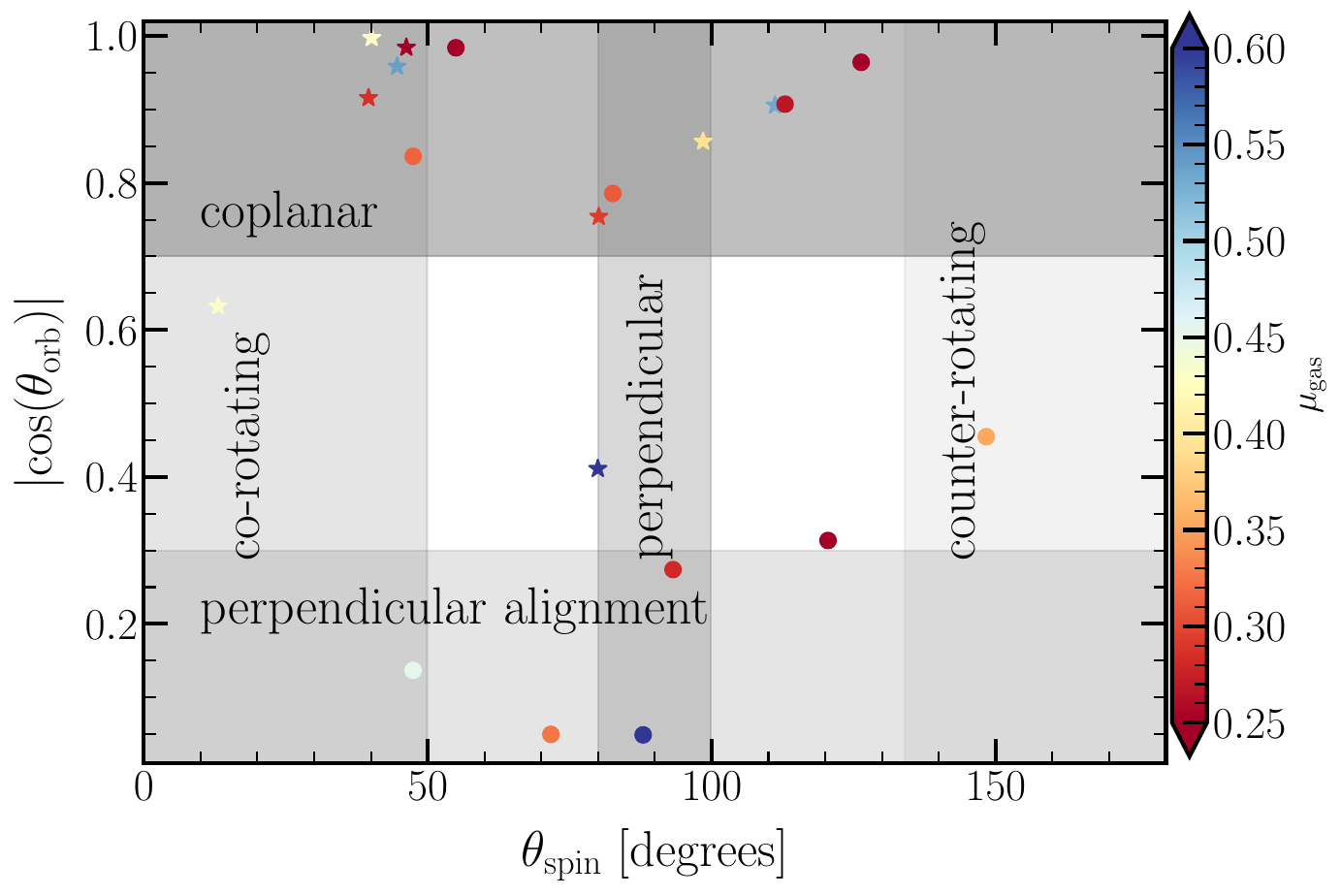} \\
  \end{tabular}
  \caption{Distribution of major mergers in the $|\cos\theta_{\rm orb}|$ versus $\cos\theta_{\rm spin}$ plane. Stars and circles correspond to BL and BD galaxies, respectively. The colour coding represents the stellar mass ratio (top panel) and the gas fraction (bottom panel). Grey hatched regions indicate the angular thresholds used to classify merger types, as defined in Fig.~\ref{fig:mergersparam}. Major mergers experienced by BL  galaxies tend to be more coplanar and corotating, and are associated with higher stellar mass ratios and gas fractions compared to those in BD galaxies.}
  \label{fig:mergersparam2}
  \end{figure}

For this section, we focus on a subsample of BL and BD galaxies that have each experienced exactly one major merger at $z<2$. This selection results in a reduced sample of $9$ BL and $12$ BD galaxies. Given the limited sample size, we do not further subdivide the galaxies by total stellar mass in this section.

We traced the evolution of several key galaxy properties: the masses of the thin disc, bulge, and stellar halo and the star formation rate, the gas fraction, the black hole mass, and specific stellar angular momentum. These quantities are examined as a function of the normalised time parameter $\eta_{\rm dyn}$, defined in Eq.~\ref{eq:etadyn}, which represents the time relative to the lookback time of the major merger in terms of the dynamical time of the halo. A detailed description of the results is provided in Appendix~\ref{app:timeMM}, and the evolution of these properties is illustrated in Fig.~\ref{fig:evolutionrelative}.

Our findings indicate that major mergers have more influence on the evolution of the thin disc, bulge, and stellar halo masses in BD galaxies compared to BL galaxies. Specifically, concerning the evolution of the bulge-to-disc mass fraction ($B/D$), BL galaxies remain bulgeless throughout the merger, while BD galaxies show an increase in $B/D$ (see Fig.~\ref{fig:evolutionrelative}). While the global star formation rates do not show strong differences between the two samples during the merger.

To understand the evolution, we now proceed to investigate the merger parameters defined in Section~\ref{subsec:mergersparam}  for BL  and BD galaxies. Figs.~\ref{fig:mergersparam:a} and \ref{fig:mergersparam:c} present the probability distribution functions of the stellar mass ratio ($\mu_{*}$) and the gas mass ratio ($\mu_{\rm gas}$), respectively. These distributions suggest that BL  galaxies typically experience mergers with higher $\mu_{*}$ and $\mu_{\rm gas}$ compared to BD galaxies. Specifically, the median stellar mass ratio for BL  galaxies is $\mu_{*} = 0.51$, whereas for BD galaxies it is $\mu_{*} = 0.39$. With an Anderson-Darling (AD) test (p-value $=0.25$), we can not rule out that the two populations are drawn from the same stellar mass ratio distributions. This difference is even more pronounced in the gas mass ratio, with medians of $\mu_{\rm gas} = 0.43$ for BL  and $0.18$ for BD galaxies. These results indicate that mergers experienced by BL  galaxies tend to be more gas-rich than those by BD galaxies. We find with an AD test (p-value$=< 0.003$), that the difference is unlikely to be drawn by the same gas mass ratio distribution.

In Fig.~\ref{fig:mergersparam:b} and \ref{fig:mergersparam:d}, we show the probability distribution functions of the orbital alignment ($|\cos \theta_{\rm orb}|$) and spin orientation ($\cos \theta_{\rm spin}$) of the merging galaxies respectively, as defined in Eq.~\ref{eq:angles}. Orbital alignments are classified as coplanar for $|\cos(\theta_{\rm orb})| > 0.7$ and as perpendicular for $|\cos(\theta_{\rm orb})| \leq 0.3$. Spin alignments are divided into corotating ($|\cos(\theta_{\rm spin})| > 0.7$, corresponding to angles between $0^\circ$–$45^\circ$), perpendicular ($-0.15 < \cos(\theta_{\rm spin}) < 0.15$, or $80^\circ$–$99^\circ$), and counterrotating ($\cos(\theta_{\rm spin}) < -0.7$, or $135^\circ$–$180^\circ$). The median orbital alignment for BL  galaxies is $|\cos \theta_{\rm orb}| = 0.90$, compared to $0.56$ for BD galaxies, indicating that mergers involving BL  galaxies are more likely to be coplanar. In contrast, BD galaxies exhibit a wider range of orbital alignments. We cannot exclude the possibility that both distributions are drawn by the same distribution with an AD test (p-value$<0.08$), which may be due to the small size of our sample. Even though we have not detected any perpendicular alignment in the mergers that our BL galaxies have undergone. For spin alignments, BL  galaxies have a median value of $\theta_{\rm spin} = 46.2^\circ$, consistent with corotation, while BD galaxies show a median of $\theta_{\rm spin} = 90.6^\circ$, indicative of predominantly perpendicular mergers. From an AD test of these distributions (p-value$<0.03$), we rule out that both samples are drawn from the same distribution.
Lastly, Fig.~\ref{fig:mergersparam2} shows orbital versus spin alignment, colour-coded by $\mu_{*}$ (top panel) and $\mu_{\rm gas}$ (bottom panel). This figure summarises our key findings. Among the BL  galaxies, seven out of nine mergers are coplanar ($|\cos \theta_{\rm orb}| > 0.7$), and   four out of nine are both coplanar and corotating. In contrast, BD galaxies display a more diverse distribution: five out of eleven are coplanar, and only one out of eleven is both coplanar and corotating. In summary, BL  galaxies tend to undergo major mergers that are more gas-rich, with higher stellar mass ratios, and more frequently involve coplanar and corotating configurations. While our sample sizes are limited, the AD tests suggest that differences in gas mass ratios and spin alignments between the two populations are statistically significant, supporting the notion that merger parameters and gas content influence the morphological evolution of galaxies\footnote{We also performed Kolmogorov–Smirnov tests, finding similar results.}.

\subsection{Star formation histories}
\label{subsec:sfh}

\begin{figure}
  \includegraphics[width=1\columnwidth]{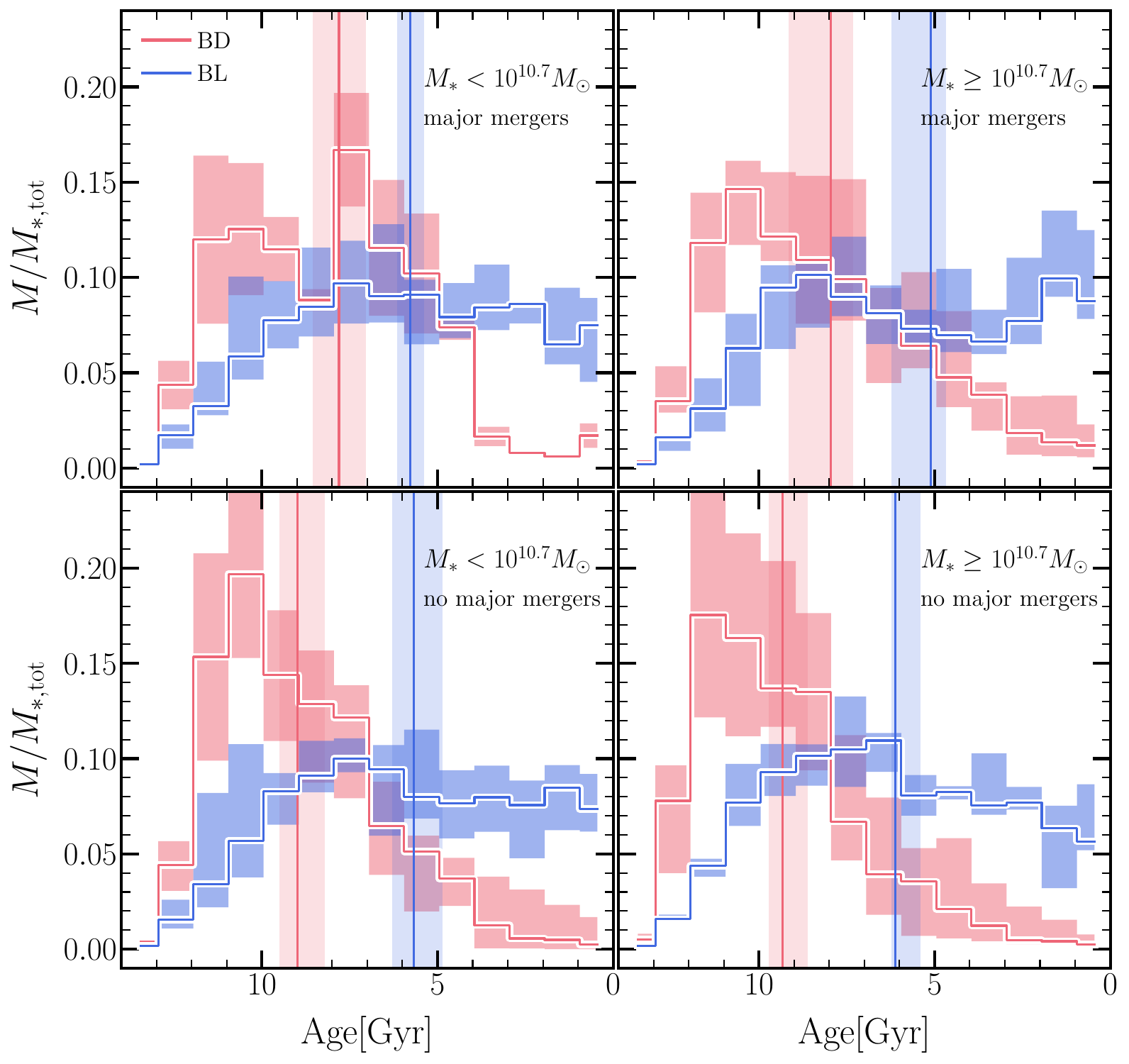}
  \includegraphics[width=1\columnwidth]{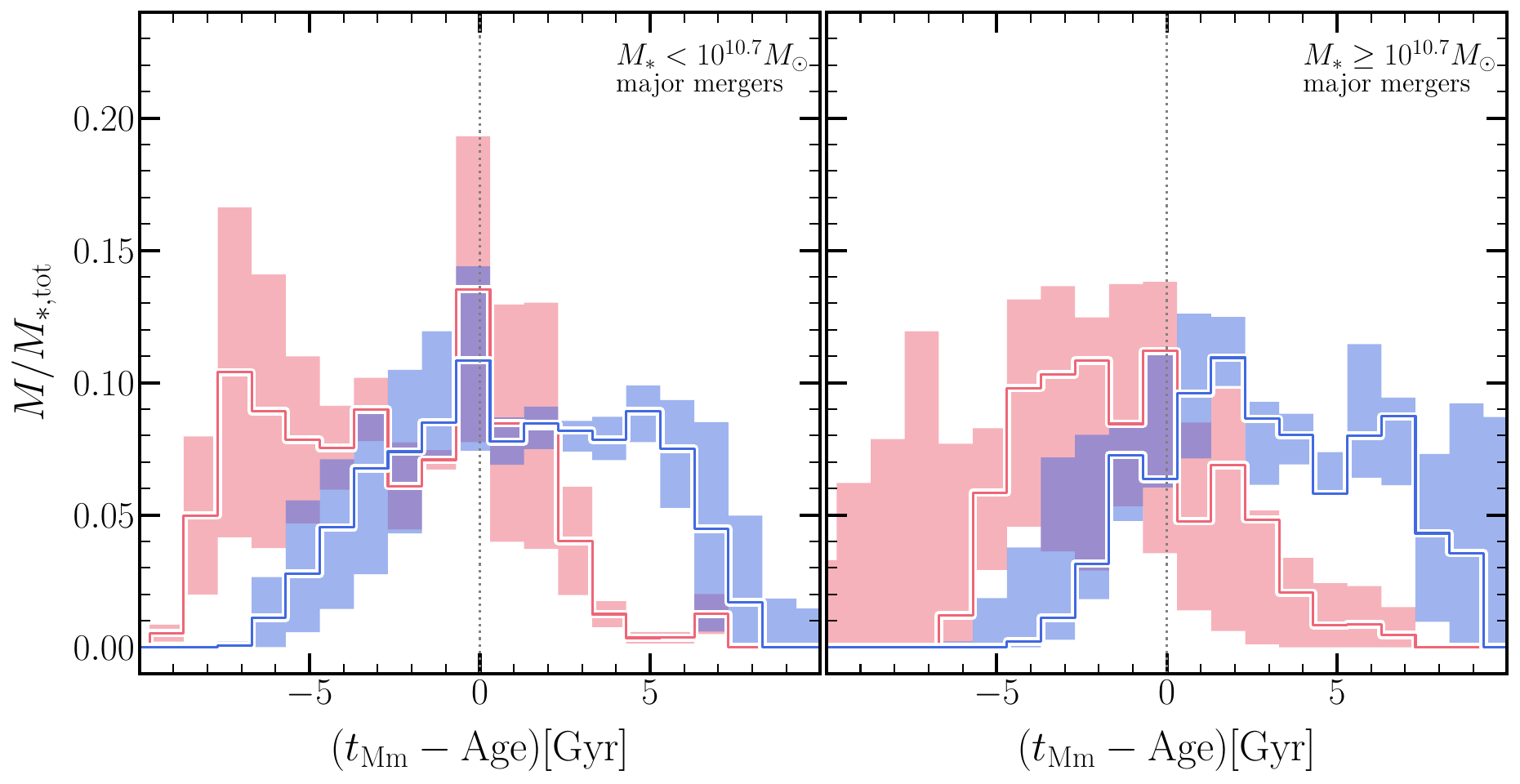}
  \caption{Star formation histories. Stacked mass-weighted age distributions of the stellar populations for BL  and BD galaxies, subdivided by stellar mass and by whether the galaxy experienced a major merger (top panel). Stacked mass-weighted age distributions relative to the time of the major merger for galaxies that have suffered just a major merger since $z=2$ (bottom panel). Shaded regions denote the $20^{\rm th}$–$80^{\rm th}$ percentile range, while vertical solid lines indicate the median star age at which $50\%$ of the $z=0$ stellar mass was assembled. BL galaxies generally host younger stellar populations compared to BD galaxies, regardless of merger history.}
  \label{fig:sfh}
  \end{figure}

Fig.~\ref{fig:sfh} shows the stacked mass-weighted age distributions of the stellar populations of the three samples. On average, BL  galaxies host younger stellar populations than BD counterparts, with a median age difference of $\sim3.2$~Gyr. This discrepancy is more pronounced in galaxies without major mergers (up to $3.3$~Gyr), and the smallest difference corresponds to low-mass galaxies that experienced a major merger ($2.0$~Gyr).

BD galaxies with mergers tend to form their stars later than BD without mergers, while BL  galaxies form at similar epochs regardless of their merger history. Star formation in BD galaxies declines rapidly over time, and these systems show little or no contribution from young stars at the present day. In contrast, BL  galaxies exhibit extended star formation histories, with a nearly flat distribution of young stars beyond the time when $50\%$ of their stellar mass has formed. An exception is found among massive BL  galaxies that have undergone a major merger, which show a localised peak in the formation of very young stars.

To further investigate the impact of mergers, we focus on galaxies that experienced a single major merger. The bottom panels of Fig.~\ref{fig:sfh} present the stacked mass-weighted age distributions relative to the last lookback  time at which both galaxies are identified for high- and low-mass systems. In low-mass galaxies, both BL  and BD types show a small enhancement in star formation within $\pm 1$~Gyr of the merger event. Notably, this peak is comparable in both populations, BL  ($M_{*}/M_{\rm tot}=0.12-0.13$) and BD galaxies ($0.11-0.12$).

\section{Discussion}
\label{sec:discussion}
  \begin{figure}
  \includegraphics[width=1.\columnwidth]{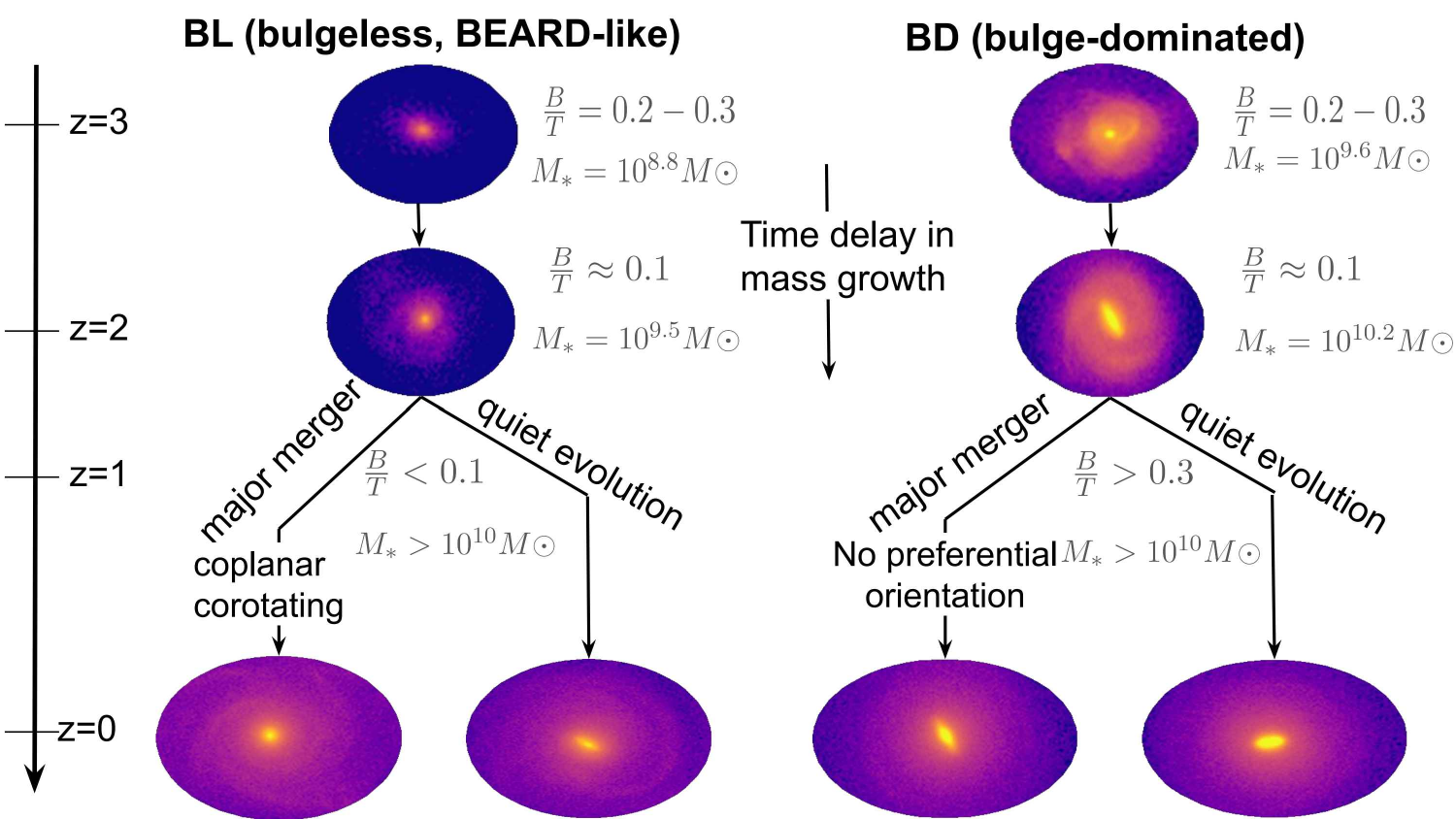}
  \caption{Schematic illustration of the evolutionary pathways of BL (bulgeless, BEARD-like) galaxies versus BD (bulge-dominated) galaxies. BL galaxies generally experience quiet merger histories, sustained growth, and smooth star formation. Although some undergo a major merger after $z=2$, these events tend to be coplanar, corotating, and gas-rich, which disrupt but do not destroy their discs. In contrast, BD galaxies assemble earlier, with significant gas accretion and active merger activity occurring primarily before $z=2$. Major mergers in BD galaxies show no preferred alignment, contributing to their bulge-dominated morphology.}
  \label{fig:summary}
  \end{figure}

In this section, we summarise the evolutionary paths of BL and BD galaxies in TNG50. We discuss our findings in the context of the MW evolution and the BEARD survey.

\subsection{Formation pathways of bulgeless (BEARD-like) and bulge-dominated galaxies}

The evolutionary channels of bulgeless (BL, BEARD-like) and bulge-dominated (BD) galaxies are summarised schematically in Fig.~\ref{fig:summary}. Within each category, we distinguish between galaxies that have undergone a major merger since $z=2$ and those that have not. BL galaxies exhibit a gradual and quiescent evolution. They accumulate $50\%$ of their stellar mass relatively late, by $z=0.7$ ($\sim6.3$ Gyr ago; Fig. \ref{fig:massevolution}). They exhibit smooth star formation histories (Fig. \ref{fig:sfh}) in agreement with observational studies of the SFHs in thin and thick discs of late-type galaxies \citep[e.g.][]{sattler2025}. Their stellar discs grow progressively over time, maintaining a dynamically cold structure while their bulges barely grow with time ($B/T=0.1$ since $z=1.5$). In contrast, BD galaxies accumulate their mass at a rapid pace. By $z=1.2$ ($\sim8.7$ Gyr ago; Fig. \ref{fig:massevolution}), they accumulate $50\%$ of their stellar mass, frequently through intense star formation (Fig. \ref{fig:sfh})  and more frequent mergers (Fig. \ref{fig:mergerhistories}) at high redshift.  Despite both populations having comparable major merger fractions since $z=2$ ($45\%$ for BL and $50\%$ for BD; Table~\ref{table:samples}), the nature of these mergers is significantly different. In BL galaxies, mergers are primarily gas-rich, coplanar, and corotating (Fig.~\ref{fig:mergersparam} and \ref{fig:mergersparam2}), leading to minimal disc disruption and allowing the disc structure to survive. This aligns with previous studies using the TNG50 simulation \citep{sotillo2022}. Conversely, BD galaxies undergo more diverse and destructive mergers, characterised by lower gas fractions and less aligned orbital parameters, contributing to substantial disc heating and bulge growth (see \citealt{wu2025}).
BD and BL galaxies can also form without recent major mergers, suggesting that the role of early gas accretion and the alignment of angular momentum of the infalling material could also play a role in morphological evolution \citep{sales2012,jiang2025}. Further investigation is required in this context.

\subsection{BL galaxies compared to the MW}
We focus on the study of bulgeless galaxies, which are now regarded as MW-like galaxies, as recent research has indicated. In particular, \cite{shen2010} propose that the MW bulge is primarily a secularly built bulge that was formed through disc instabilities from a pure disc galaxy. The bulge contribution is estimated to be at most $8\%$ of the disc mass, and \cite{shen2010} can replicate the observations of our galaxy. Furthermore, the GAIA mission has shown that the MW underwent a major merger, specifically Gaia-Sausage-Enceladus, which was the most recent massive merger approximately 10 Gyr ago \citep{helmi2018,belokurov2018}. It did not destroy the MW disc and triggered star formation, and the estimated stellar mass ratio of this merger is $1:4$ ($\mu_{*}={0.25}$). Our analysis indicates that $74\%$ of the galaxies experienced at least one significant merger during their evolution (Fig.~\ref{fig:mergerhistories}). Nevertheless, the orbital angles of the present stellar discs (which are primarily coplanar) and their aligned rotation (Fig.~\ref{fig:mergersparam2}) prevent this event from destroying them. The remaining $26\%$ of BL galaxies exhibit a calm evolution, with no major mergers.

Bulgeless galaxies are likely to be located in lower-mass dark matter haloes \citep{marrero2025}, exhibit higher specific star formation rates, and have higher hydrogen gas fractions than their BD galaxy counterparts for a given stellar mass (Fig. \ref{fig:galaxyproperties}). This is in alignment with previous research \citep{sotillo2022,proctor2024,rodriguez2025}. Indeed, comparing the properties of the BL galaxies with analytic models that match observations from different surveys focused on the Milky Way such as GAIA DR2, VERA and APOGEE, we find that the median $\rm log_{10}(M_{*}/M_{200})_{\rm BL }=-1.30$ and  $\rm log_{10}(j_{*,\rm BL })=3.16 \,\rm kpc\,\rm km\, s^{-1}$,  are close to the estimates from the MW ($\rm log_{10}(M_{*}/M_{200})_{MW}=-1.33$; \citealt{cautun2020}, $\rm log_{10}(j_{*,\rm MW})=3.03 \,\rm kpc\,\rm km\, s^{-1}$;
\citealt{obreja2022}). In contrast, the median  $\rm sSFR_{\rm BL }=9.77\times10^{-11}\,\rm yr^{-1}$ and $\log_{10}(f_{\rm H\textsc{i}+H_{2}}) = -0.43$ are higher in BL galaxies than the MW estimates  ($\rm sSFR_{\rm MW}=2.71\pm0.59\times10^{-11}\,\rm yr^{-1}$; \citealt{licquia2015} and $\log_{10}(f_{\rm H\textsc{i}+H_{2}}) = -0.67$; \citealt{mcmillan2017}). Although one might could find that the early star formation of BD galaxies resembles to the MW  early high star-formation phase (e.g. \citealt{xiang2025}), such a comparison is not straightforward given that our results describe galaxy population-level trends, whereas the MW  benefits from detailed reconstruction using Gaia and LAMOST data. For this reason, we do not interpret the MW early disc formation at this stage as a direct analogue of BD or BL galaxies. However, in order to accurately determine whether the assembly history of the MW is an outlier or the norm, a more equitable comparison with significant statistics MW-analogues is required. BEARD (M{\'e}ndez-Abreu et al. in prep) was developed in response to this need. A detailed comparison could be carried out in the following studies.

\subsection{Some caveats regarding simulations}
This work explores the role of mergers in the evolutionary pathways of BL and BD galaxies in the TNG50 simulation, rather than comparing them to BEARD observations.
TNG50 provides a cosmological framework for researching galaxy evolution, including bulgeless galaxies, although it has limits. Although subgrid models can reproduce large-scale observables, they may not fully resolve the physical mechanisms that regulate central mass building, including feedback processes \citep[][see their Fig. A1]{tacchella2019}. The evolutionary trends found are nonetheless important despite these limitations. Instead, they highlight the need for complementary approaches, including detailed observations such as BEARD, to further refine our understanding of the origin and frequency of MW galaxies in a cosmological context.

\section{Summary}
\label{sec:summary}
In this paper, we have investigated the formation and evolution of bulgeless (BL) galaxies using the M31/MW analogues catalogue from TNG50 \citep{pillepich2023}. This catalogue includes galaxies with a $M_{\rm halo} (M_{*})<10^{13} (10^{10.5-11.2}) \, \Msun$, visually disky or with a minor-to-major axes smaller than $0.45$ and isolation. We apply a morphological selection that aligns with Bulge-to-Disc mass fractions $B/D\leq 0.08$, comparable to the MW fraction \citep{shen2010} and BEARD survey. For the last condition, we identify distinct kinematic components using \textsc{Mordor} \citep{zana2022}. For comparison, we chose a sample of galaxies having a dominant bulge (BD) of $B/D>1$. The remaining sample consists of $31$ BL and $32$ BD galaxies from the parent sample (PS)
Our main findings are the following:

\begin{itemize}

\item $74\%$ of BL galaxies have undergone a major merger (stellar mass ratio of $1:4$) during their lifetime (Fig. \ref{fig:mergerhistories}). This is in agreement with \cite{sotillo2022} on the same simulation dataset. In addition, BD galaxies have more active major merger histories than BL galaxies, with $91\%$ of BD galaxies undergoing a major merger during their lifetime. Nevertheless, approximately half of both the BD and BL samples undergo at least a major merger since $z=2$.

\item On average, BL galaxies form $50\%$ of their final stellar mass $6.3$ Gyrs ago ($z_{50}=0.7$) and form later than BD galaxies ($7.2$ Gyrs ago, ie., $z_{50}=1.2$; Fig. \ref{fig:massevolution}).

\item  At a fixed stellar mass, BL galaxies are hosted in lower-mass dark matter haloes, exhibit higher specific angular momentum, specific star formation rates and hydrogen gas fractions than BD galaxies (Fig. \ref{fig:galaxyproperties}), in alignment with previous studies \citep{sotillo2022,proctor2024,rodriguez2025}. The properties of BL galaxies are compatible with MW estimates from GAIA DR2, VERA and APOGEE.
 %We analysed the redshift evolution of thin disc and bulge components in BL, BD, and parent sample galaxies, considering differences in stellar mass and merger history (Fig. \ref{fig:evolutionall}).

\item BL galaxies exhibit gradual disc growth and modest bulge assembly, maintaining disc-dominated morphologies (with $D/T > 0.5$ and $B/D\leq0.08$ at $z=0$, Fig. \ref{fig:evolutionall}). In contrast, BD galaxies build their discs earlier, but they experience more morphological evolution, especially in high-mass systems with major mergers, resulting in disc mass decrease (Fig.~\ref{fig:evolutionrelative}). BD galaxies also show substantial late-time bulge growth, leading to high $B/T$ ratios ($\sim0.35$–$0.45$) and $B/D>1$ by $z=0$. These divergent pathways appear as early as $z \sim 4$, suggesting not only major mergers but gas accretion and minor mergers could play a role \citep{sales2012}.

\item  In comparison to BD systems, which exhibit a broader spectrum of orbital and spin configurations, BL galaxies undergo more gas-rich, coplanar, and corotating mergers, despite similar merger frequencies (Figs. \ref{fig:mergersparam} and \ref{fig:mergersparam2}). This could elucidate the reason why BL galaxies do not undergo morphological evolution during the major merger, whereas BD galaxies do (Fig.~\ref{fig:evolutionrelative}).

\item BL galaxies have younger star populations than BD galaxies, with a difference of up to $3.3$ Gyr in non-merged galaxies. BL galaxies have more extended star formation histories than BD galaxies, which have a rapid decline in star formation without young stellar populations at $z=0$. A single major merger can lead to minor star formation enhancements in both BL and BD galaxies, especially in low-mass systems (Fig. \ref{fig:sfh}). These enhancements do not significantly affect their star formation histories.

\end{itemize}

These findings show the different pathways that lead to the formation and evolution of bulgeless galaxies and bulge-dominated galaxies at $z=0$. The early assembly of bulge-dominated galaxies and high-redshift physical processes seem to play a role in accounting for the difference in the evolution. Major mergers in bulgeless galaxies can perturb the thin disc, but it later recovers, whereas in bulge-dominated galaxies, major mergers can strongly perturb the thin disc. This is caused by the initial conditions of the mergers: bulgeless galaxies are more likely to undergo coplanar, corotating, and richer-gas mergers than dominated-bulge galaxies. In a companion paper \citep[][]{cardona2026}, we also see differences in the satellite galaxy population of BL and BD galaxies, pointing also towards a preferential orbital and spin alignment of satellites in BL galaxies. This emphasises the potential of the current generations of cosmological hydrodynamical simulations and the multi-observational programme BEARD (M{\'e}ndez-Abreu et al. in prep), which facilitates a thorough understanding of the history of bulgeless galaxies, including our own galaxy, the Milky Way.

  \begin{acknowledgements}
We thank the anonymous referee for their constructive comments, which improved the clarity of the paper. YRG acknowledges finacial support from Plan Propio de Investigaci\'on 2025 submodalidad 2.3 of the University of Cordoba.  AdLC  acknowledges financial support from the Spanish Ministry of Science and Innovation (MICINN) through RYC2022-035838-I and PID2021-128131NB-I00 (CoBEARD project). JMA acknowledges the support of the Agencia Estatal de Investigación
del Ministerio de Ciencia e Innovación (MCIN/AEI/10.13039/501100011033) under grant nos. PID2021-128131NB-I00 and CNS2022-135482 and the European Regional Development Fund (ERDF) ‘A way of making Europe’ and the ‘NextGenerationEU/PRTR'. EAG acknowledges support from the Agencia Espacial de Investigación del Ministerio de Ciencia e Innovación (AEI-MICIN) and the European Social Fund (ESF+) through a FPI grant PRE2020-096361. ADC thanks the Ministry of Science and Innovation for funding though the 2023 Consolidación Investigadora program (grant code CNS2023-144669). MCC acknowledges the support of AC3, a project funded by the European Union's Horizon Europe Research and Innovation programme under grant agreement No 101093129. MCC acknowledges financial support from the Spanish Ministerio de Ciencia, Innovación y Universidades (MCIU) under the grant PID2021-123417OB-I00 and PID2022-138621NB-I00.
SZ acknowledges the financial support provided by the Governments of Spain and Arag\'on through their general budgets and the Fondo de Inversiones de Teruel, the Aragonese Government through the Research Group E16\_23R, and the Spanish Ministry of Science and Innovation and the European Union-NextGenerationEU through the Recovery and Resilience Facility project ICTS- MRR-2021-03-CEFCA. FP acknowledges support from the Horizon Europe research and innovation programme under the Maria Skłodowska-Curie grant “TraNSLate” No 101108180, and from the Agencia Estatal de Investigación del Ministerio de Ciencia e Innovación (MCIN/AEI/10.13039/501100011033) under grant (PID2021-128131NB-I00) and the European Regional Development Fund (ERDF) ``A way of making Europe''. EMC and AP acknowledge the support by the Italian Ministry for Education University and Research (MUR) grant PRIN 2022 2022383WFT “SUNRISE", CUP C53D23000850006 and Padua University grants DOR 2022-2024.
SCB acknowledges the support of the Agencia Estatal de Investigación del Ministerio de Ciencia e Innovación (MCIN/AEI/10.13039/501100011033) under grant nos. PID2021-128131NB-I00 and CNS2022-135482 and the European Regional Development Fund (ERDF) ‘A way of making Europe’ and the ‘NextGenerationEU/PRTR’. The project that gave rise to these results received the support of a fellowship from the “la Caixa” Foundation (ID 100010434). The fellowship code is LCF/BQ/PR24/12050015. LC acknowledges support from grant PID2022-139567NB-I00 funded by the Spanish Ministry of Science and Innovation/State Agency of Research MCIN/AEI/10.13039/501100011033 and by “ERDF A way of making Europe”.VC is supported by ANID through the FONDECYT grant nr 11250723.

  \end{acknowledgements}


\appendix

\section{The TNG50 simulation}
\label{app:TNG50sim}
 We focus on the \TNGF~simulation \citep{pillepich2019,nelson2019b}, which is the highest resolution simulation that is part of the TNG project \footnote{\citep{nelson2019a}; http://www.tng-project.org}.
 %(\citealt{nelson2018,naiman2018,pillepich2018b,marinacci2018,springel2018}).
 The simulation evolves $2160^3$ dark matter particles and initial gas cells in a 51.7 comoving Mpc region from $z=127$ down to $z=0$. The mass resolution is $4.5\times 10^5 \Msun$ for dark matter particles, whereas the mean gas mass  cell resolution is $8.5\times10^4\Msun$. A comparable initial mass is passed down to stellar particles, which subsequently lose mass through stellar evolution.
  The spatial resolution for collisionless particles (dark matter, stellar, and wind particles) is 575 comoving pc down to $z=1$, after which it remains constant at 288 pc in physical units down to $z=0$. In the case of the gas component, the gravitational softening  is adaptive and based on the effective cell radius,
  down to a minimum value of 72 pc in physical units, which is imposed at all times.

  Galaxies and their haloes are identified as bound substructures using an \textsc{FoF} (Friend-of-Friends) and then a \textsc{SUBFIND} algorithm \citep{springel2001}.
  Halo masses ($M_{200}$) are defined as all matter within the radius $R_{200}$ for which
  the inner mean density is $200$ times the critical density. The most massive subhalo in each FoF group is considered the central galaxy, and the remaining ones are satellites. Stellar mass is computed as the total stellar content bound to each subhalo.

\section{Kinematic decomposition}
\label{ap:mordor}
  We use the dynamical decomposition from \cite{zana2022} who uses \textsc{mordor} code\footnote{ \url{https://github.com/thanatom/mordor}} to determine more specific galaxy components in Section \ref{sub:discsample}. The decomposition is based on the circularity ($\epsilon$) and binding energy ($E$) phase space, where a minimum in $E$ is identified for each galaxy, $E_{\rm cut}$. A maximum of four dynamical components can be identified:
  \begin{itemize}
      \item The classical Bulge comprises stellar particles that exhibit the highest degree of binding ($E\leq E_{\rm cut}$), whose value depends on the $E_{\rm cut}$ of each galaxy and exhibit counterrotation, characterised by negative values of $\epsilon$ ($\epsilon<0.0$). Using Monte Carlo sampling, an equal distribution of corotating stellar particles is chosen and allocated to the bulge.
      \item The thin or cold disc: is made up of stellar particles that exhibit a high degree of binding and are not assigned to the bulge but have positive values of $\epsilon$ ($\epsilon>0.7$).
      \item The secular bulge or pseudobulge, is formed by the remaining stellar particles that exhibit a high degree of binding, but they are not assigned to the bulge, such as $0<\epsilon<0.7$ and $E\leq E_{\rm cut}$.
      \item The stellar halo comprises stellar particles that have a lower bound ($E >E_{\rm cut}$) and are assigned to the stellar halo with the same methodology as for the bulge.
      \item The thick or warm disc  is composed of the stellar particles that exhibit a lower degree of binding ($E >E_{\rm cut}$) but with values of $0<\epsilon<0.7$.

  \end{itemize}

\section{Merger histories}
\label{ap:mh}
\begin{figure}
      \centering
      \includegraphics[width=1\linewidth]{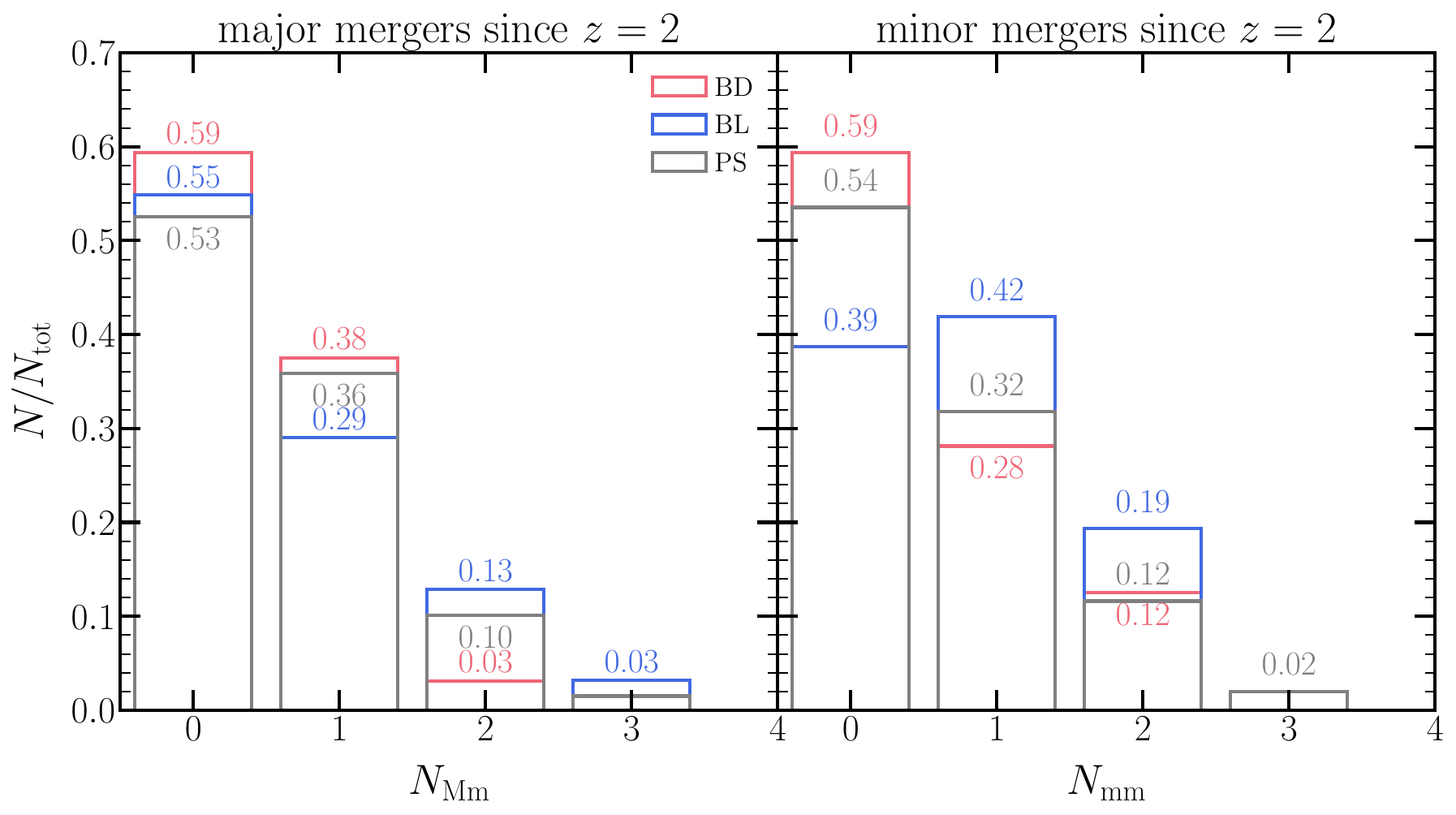}

      \caption{Merger histories of bulgeless (BL)  and BD galaxies since $z=2$. The left and right panels respectively depict the fraction of BL, BD, and PS galaxies as a function of the number of major and minor mergers. The merger histories of the samples exhibit a slight difference of no more than $4\%$, as approximately half of the galaxies in each sample have undergone at least one major merger since $z=2$. The disparity is more pronounced in the case of minor mergers, with at least one minor merger occurring in $61\%$ of the BL  galaxies and $41\%$ of the BD galaxies.}
      \label{fig:mergerhistoryz2}
  \end{figure}

In this appendix,  we show the differences in the merger histories of the BL galaxies and compare them with BD and PS galaxies described in Section \ref{sub:discsample} and mergers considered since $z=2$.
In Fig. \ref{fig:mergerhistoryz2}, we summarise our findings: BL and BD galaxies have comparable major merger histories with approximately half of the galaxies in each sample having undergone a major merger since $z=2$, with a galaxy fraction of $0.59$ for BD and $0.55$ for BL galaxies. On the contrary, minor merger histories have indicated larger disparities among samples. BL galaxies present a galaxy fraction of $0.39$ without minor mergers, while BD galaxies exhibit the highest fraction ($61$ per cent).

\section{Evolution of all dynamical components in BL and BD}
\label{app:morphev}

\begin{figure}
  \includegraphics[width=\columnwidth]{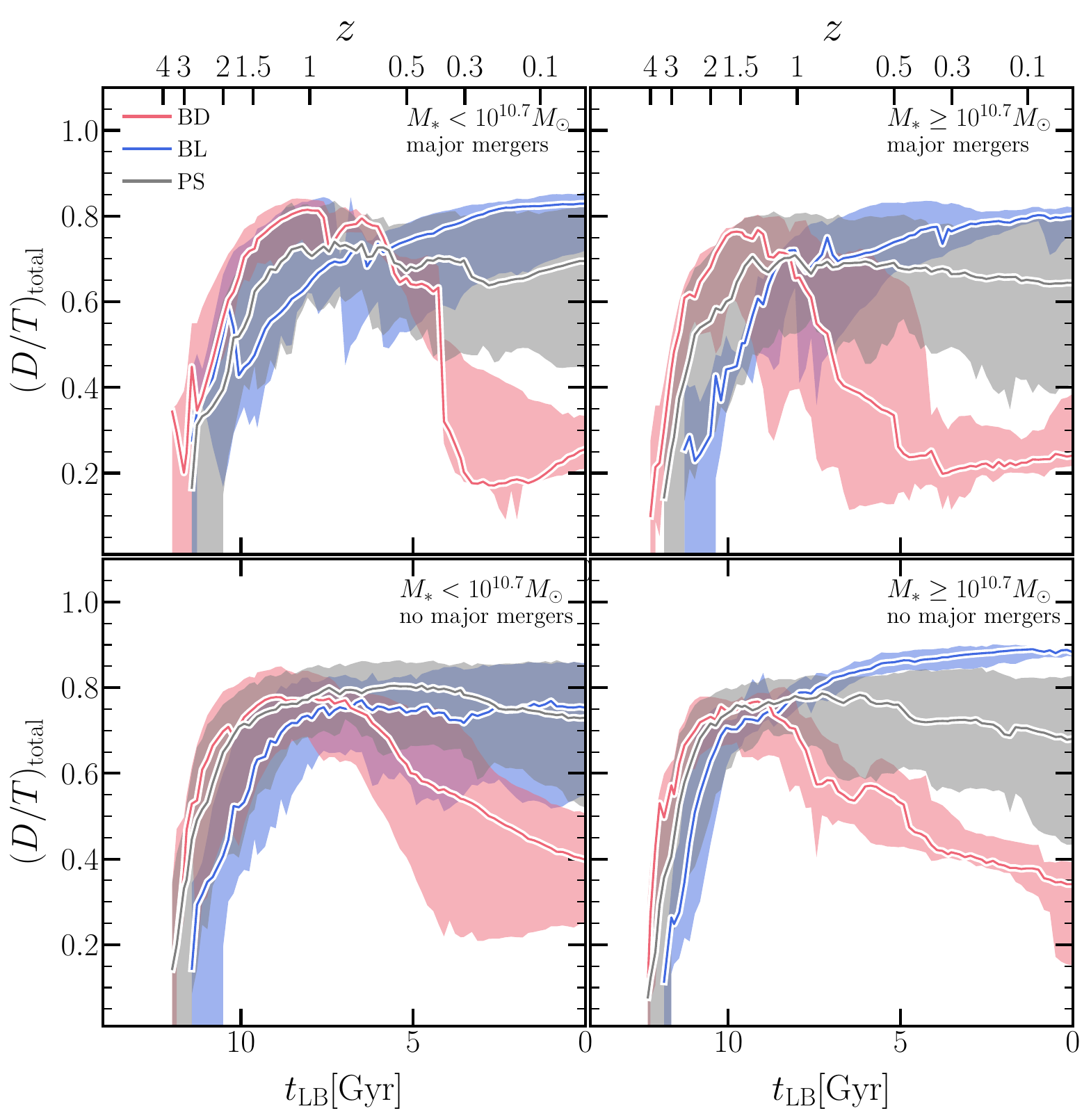}
  \caption{Evolution of total disc-to-total mass fraction, encompassing the thin, thick, and secular bulge components across the three samples.  The total disc-to-total mass fraction evolution shows a comparable trend to that depicted in Fig.~\ref{fig:evolutionallc}, except for higher values of $(D/T)_{\rm total}$, surpassing $0.7$ in BL galaxies at $z=0$.}
  \label{fig:evolutionDtoTtot}
  \end{figure}

This appendix presents the evolution of the total disc-to-total mass fraction, $(D/T)_{\rm total}$, for BL, BD and PS samples. The total disc-to-total mass fraction comprises the three rotating components (thin and thick disc and secular bulge) identified by dynamical decomposition. Fig~\ref{fig:evolutionDtoTtot} illustrates the evolution of the disc-to-total mass fraction. Upon comparing the evolution of $(D/T)_{\rm total}$ with that of the thin disc-to-total mass fraction (see \ref{fig:evolutionallc}),  we observe analogous evolutions while preserving differences between the samples. Nevertheless, the total disc-to-total mass fractions ($>0.7$) are systematically higher than the thin disc-to-total mass fraction for the BL galaxies.

\section{Galaxy properties evolution relative to the lookback time of the major merger}
\label{app:timeMM}
  \begin{figure*}
  \includegraphics[width=2\columnwidth]{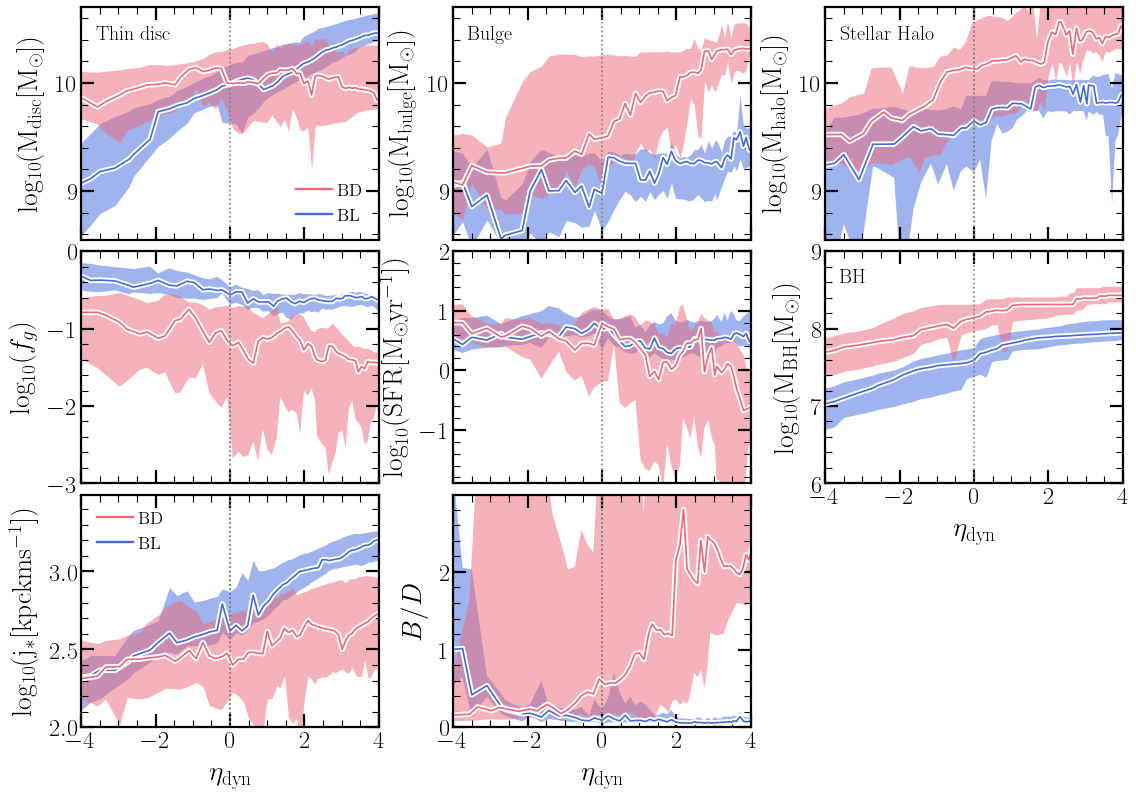}
  \caption{Evolution of galaxy properties as a function of $\eta_{\rm dyn}$ defined in Eq. \ref{eq:etadyn}. Top panels: Thin disc, bulge, and stellar halo mass evolution.  Middle panels: Gas fractions, star formation rates within twice the half stellar mass radius, and the black hole growth. Bottom panels: Specific angular momentum and $B/D$. Vertical dotted lines correspond to the lookback time at which the major merger happens. Major mergers in BL galaxies, which appear to be BL at the time of the merger, seem to be less disruptive to the galaxy compared to BD galaxies.}
  \label{fig:evolutionrelative}
  \end{figure*}

  To better understand the structural evolution of galaxies following major mergers, we examine a subsample of systems that have experienced exactly one major merger since $z = 2$. This selection yields 9 BL  and 12 BD galaxies. Due to the small sample size, we do not further divide the galaxies by stellar mass.
  We define the normalised dynamical time parameter, $\eta_{\rm dyn}$, as:

  \begin{equation}
    \eta_{\rm dyn} =  (t_{\rm MM}- t_{\rm LB})/t_{\rm dyn},
  \label{eq:etadyn}
  \end{equation}
  where $t_{\rm LB}$ is the lookback time, $t_{\rm MM}$ is the lookback time of the major merger, and $t_{\rm dyn}$ is the halo dynamical time, defined as the free-fall time of a dark matter halo:
  \begin{equation}
    t_{\rm dyn}\equiv \left( \frac{ 3\pi}{32G(200 \rho_{\rm crit})}\right)^{1/2}.
  \end{equation}
  The value of $t_{\rm dyn}$ depends on redshift, with approximate values of 0.9 Gyr at $z = 1$, 1.2 Gyr at $z = 0.5$, and 1.6 Gyr at $z = 0$. Positive $\eta_{\rm dyn}$ indicates times after the merger; negative values indicate times before.

  The evolution of the thin disc, bulge, and stellar halo masses in BL  galaxies as a function of $\eta_{\rm dyn}$ is illustrated in the upper panels of Fig.~\ref{fig:evolutionrelative} from left to right. It is evident that the thin disc experienced some disruption in less than two dynamical times; however, it was able to recover and grow over time. Similarly, the bulge and stellar halo experience a slight increase in mass, just above $50\%$, after two dynamical times.   For comparison, we have also incorporated the evolution of BD galaxies, which exhibit more extreme behaviour. After two dynamical times, the thin disc is disrupted, resulting in a decrease in average mass of $20\%$. This decrease occurs continuously. Before the main merger occurred, the bulge and stellar halo masses began to increase by approximately $32\%$ and $50\%$, respectively, in one dynamical time after the merger took place. Nevertheless, the evolution of the components of the BD galaxies exhibits a greater degree of scatter. In fact, following the bulge-to-disc mass fraction ($B/D$, see bottom panel of the middle column), BL galaxies remain bulgeless during the merger, whereas BD galaxies exhibit an increase in $B/D$.

  The gas fraction ($f_{\rm g}=M_{\rm gas}/ (M_{*}+M_{\rm gas})$), the star formation rate within an aperture of twice the stellar half mass radius as a function of $\eta_{\rm dyn}$ and the growth of black hole mass are depicted in the middle panels of the figure from left to right. Bulgeless galaxies present a gradual and minor decrease in the gas fraction and the star formation rate after the major merger happens. Additionally, the black hole mass grows gradually subsequent to the major merger. The bottom panel of Fig.~\ref{fig:evolutionrelative} illustrates the evolution of the specific angular momentum, which exhibits minor fluctuations in a dynamical time period prior to and following the merger, followed by a recovery and subsequent increase over time.
  In comparison, the larger bulge galaxies experience a more significant decline in gas fraction and star formation rates subsequent to the main merger. As time progresses, the mass of the black hole increases.  The stellar specific angular momentum does not exhibit evolution in a single dynamical time before and after the merger; rather, it increases over time at a later time.  We also observe that the primary differences in the evolution of galaxy properties between BL  and BD galaxies begin earlier, which implies that other physical processes, such as the type of gas inflows falling into the halo, occur at higher redshifts in addition to major mergers. The parameters of the major mergers have a distinct impact on BL  and BD galaxies.

  \end{document}